\documentclass[onecolumn,showpacs,amsmath,amssymb,superscriptaddress,footinbib]{revtex4}

\usepackage{amsfonts}
\usepackage{graphicx}
\usepackage{color}

\def\hmpc{\ {\rm {\it h}^{-1}Mpc}}

\def\hmsun{\ {\rm M_\odot/{\it h}}}

\def\hmmpc{\ {\rm {\it h}Mpc^{-1}}}

\def\la{\langle}
\def\ra{\rangle}

\def\ln{{\rm ln}}

\def\zvh{\mathrm{\hat{\bf{z}}}}

\def\vk{\mathrm{\bf k}}

\def\vx{\mathrm{\bf x}}
\def\vy{\mathrm{\bf y}}

\def\bias#1{{\rm b}_{\rm{#1}}}

\def\xnu{\xi_{>\nu}}
\def\pnu{P_{>\nu}}

\def\fnl{f_{\rm NL}}
\def\gnl{g_{\rm NL}}

\newcommand{\be}{\begin{equation}}
\newcommand{\ee}{\end{equation}}
\newcommand{\bea}{\begin{eqnarray}}
\newcommand{\eea}{\end{eqnarray}}
\def\ba#1\ea{\begin{align}#1\end{align}}
\newcommand{\D}{\Delta}
\newcommand{\rhob}{\overline{\rho}}
\newcommand{\Om}{\Omega_m}

\newcommand{\s}{\sigma}
\renewcommand{\d}{\delta}
\renewcommand{\a}{\alpha}
\renewcommand{\o}{\omega}
\newcommand{\Sk}{S^{(3)}}
\newcommand{\eps}{\varepsilon}

\newcommand{\wt}{\omega^{(2)}}
\newcommand{\wth}{\omega^{(3)}}

\renewcommand{\v}[1]{\bm{#1}}
\newcommand{\<}{\langle}
\renewcommand{\>}{\rangle}
\newcommand{\bo}{{\rm b}_{\rm I}}
\newcommand{\bt}{{\rm b}_{\rm II}}
\newcommand{\nh}{\bar{n}_h}
\newcommand{\nhh}{\hat{\bar{n}}_h}
\newcommand{\Mpch}{\,{\rm Mpc}/h}
\newcommand{\iMpch}{\,h/{\rm Mpc}}

\newcommand{\refeq}[1]{Eq.~(\ref{eq:#1})}          
\newcommand{\refeqs}[2]{Eqs.~(\ref{eq:#1})--(\ref{eq:#2})}          
\newcommand{\refe}[1]{(\ref{eq:#1})}          
          
\newcommand{\reffig}[1]{Fig.~\ref{fig:#1}}          

\newcommand{\refsec}[1]{Sec.~\ref{sec:#1}}
\newcommand{\vs}{\nonumber\\}

\def\bm#1{\textbf{\em #1}}
\def\Mm{\mathcal{M}}

\def\erfc{\mathrm{erfc}}
\def\Pp{P_\phi}

\definecolor{RedWine}{rgb}{0.743,0,0}
\definecolor{RoyalBlue}{rgb}{0.25,.41,.88}
\definecolor{ForestGreen}{rgb}{.13,.54,.13}
\definecolor{DeepPurple}{rgb}{.72,.18,1}

\begin{document}

\title{
Non-Gaussian Halo Bias Re-examined:
Mass-dependent Amplitude from the Peak-Background Split and Thresholding} 
\author{Vincent Desjacques}
\email{dvince@physik.uzh.ch}
\affiliation{Institute for Theoretical Physics, University of Z\"urich,
Winterthurerstrasse 190, CH-8057 Z\"urich, Switzerland}
\author{Donghui Jeong}
\email{djeong@tapir.caltech.edu}
\affiliation{
California Institute of Technology, Mail Code 350-17, Pasadena, CA, 91125
}
\author{Fabian Schmidt}
\email{fabians@tapir.caltech.edu}
\affiliation{
California Institute of Technology, Mail Code 350-17, Pasadena, CA, 91125
}
\date{\today}


\begin{abstract}
Recent results of N-body simulations have shown that current theoretical 
models are not able to correctly predict the amplitude of the 
scale-dependent halo bias induced by primordial non-Gaussianity, for 
models going beyond the simplest, local quadratic case. 
Motivated by these discrepancies, we carefully examine three 
theoretical approaches based on 
(1) the statistics of thresholded regions,
(2) a peak-background split method based on separation of scales, and 
(3) a peak-background split method using the conditional mass function. 
We first demonstrate that the statistics of thresholded regions, 
which is shown to be equivalent at leading order to a local bias expansion, 
cannot explain the 
mass-dependent deviation between theory and N-body simulations.  
In the two formulations of the peak-background split on the other hand,
we identify an important, but previously overlooked,
correction to the non-Gaussian bias that strongly depends on halo mass.  
This new term is in general significant for 
any primordial non-Gaussianity going beyond the simplest
local $\fnl$ model.  
In a separate paper \cite{2011PhRvD..84f1301D}, we compare these new 
theoretical predictions with N-body simulations, showing good agreement 
for all simulated types of non-Gaussianity.  
\end{abstract}

\pacs{98.80.-k,~98.65.-r,~98.80.Cq,~95.36.+x}

\maketitle

\setcounter{footnote}{0}


\section{Introduction}
\label{sec:intro}

Ongoing and future galaxy surveys will provide a large amount of data 
that can be exploited to constrain the physics of inflation and the
very early Universe, in particular through a measurement of the shape 
and amplitude of primordial non-Gaussianity (NG).  
Over the past few years, galaxy clustering has emerged 
as the most powerful large-scale structure probe of primordial 
non-Gaussianity (e.g., \cite{2004PhRvD..69j3513S,2006PhRvD..74b3522S,
2008PhRvD..77l3514D}; for a review, see \cite{2010arXiv1006.4763D}). 
In particular, references 
\cite{2008PhRvD..77l3514D,2008ApJ...677L..77M,2008JCAP...08..031S,2008PhRvD..78l3507A} 
have shown that the local quadratic coupling 
$\fnl\phi^2$ induces a scale-dependent bias 
\begin{equation}
\D\bias{I}(k,z) = \frac{2\fnl(b_1^{\rm E}-1)\delta_c}{{\cal M}(k,z)}
\label{eq:one}
\end{equation}
in the large-scale power spectrum of biased tracers. 
Here, $b_1^{\rm E}$ is the (Eulerian) linear bias factor, 
$\delta_c\approx 1.69$ is the linear critical density contrast 
for spherical collapse, and ${\cal M}(k,z) \propto D(z) k^2 T(k)$ 
is the transfer function between density and 
the gravitational (Bardeen) potential perturbations. 
Numerical studies have confirmed the scaling $\D\bias{I}\propto k^{-2}$ 
and the redshift-dependence $\D\bias{I}\propto D(z)^{-1}$
\cite{2008PhRvD..77l3514D,2009MNRAS.396...85D,2010MNRAS.402..191P,
2009MNRAS.398..321G,2010JCAP...07..002N,2010arXiv1010.0055S}, even though 
the exact amplitude of the effect remains somewhat debatable
(at the $\sim 10-20 \%$ level; presumably related to the choice of halo 
finder \cite{2010arXiv1006.4763D}).  

However, for other non-Gaussian models such as a local $\fnl\phi^2$ model
with $k$-dependent $\fnl$, the local cubic coupling $\gnl\phi^3$, 
or the orthogonal type, there is a much larger discrepancy between the 
analytical predictions based on the statistics of high-threshold 
regions \cite{2008ApJ...677L..77M} and
the non-Gaussian bias measured from simulations
\cite{2010PhRvD..81b3006D,2010arXiv1010.3722S,2011arXiv1102.3229W}. 
In the $\gnl \phi^3$ case, the magnitude of the non-Gaussian scale-dependent 
bias $\D\bias{I}$ is significantly suppressed relative to the 
theoretical expectation on large scales ($k\lesssim 0.01\hmmpc$), 
even for highly biased halos. The ratio of the measured to the 
predicted non-Gaussian bias strongly depends on the halo mass $M$: 
it decreases towards low mass halos, and even reverses sign for 
halos with $b_1^{\rm E}\lesssim 2$ \citep{2010PhRvD..81b3006D}. 
For the quadratic coupling with $k$-dependent $\fnl(k)\propto k^{n_f}$, 
the discrepancy between the simulated bias and the high-peak
expectation also becomes more severe as the halo mass decreases. 
Furthermore, the deviation depends on the sign and amplitude of the 
spectral index $n_f$ \cite{2010arXiv1010.3722S}.  
Recent numerical simulations implementing the orthogonal bispectrum shape 
also show systematic deviations in the measured halo bias from the 
high-peak expectation, in a way that the deviation becomes larger towards
lower halo masses \cite{2011arXiv1102.3229W}.  

In this paper, we present a careful (re-)derivation of the effect of 
local and non-local primordial non-Gaussianity 
on the large-scale clustering of tracers 
(such as galaxies and clusters of galaxies) 
using the thresholding approach 
\cite{GrinsteinWise,1986ApJ...310L..21M,2008ApJ...677L..77M}, 
as well as two distinct albeit related formulations of 
the peak-background split (PBS).  For all three approaches, we present 
general expressions for the non-Gaussian scale-dependent bias and
apply them to models for which N-body simulations have been performed.  

In the thresholding approach, we directly calculate the two-point
correlation function of halos from the probability of finding 
a single smoothed region above some threshold, and the probability 
of finding two separate regions above the same threshold.  
This approach has the advantage that the thresholding process is 
a well-defined mathematical operation so that we can in principle calculate 
the correlation functions without any further approximation.  
We derive a general expression for the amplitude of the non-Gaussian bias 
in terms of the primordial $N$-point functions and the Gaussian bias 
parameters of the thresholded regions,
without relying on the high-peak assumption usually assumed in
previous studies (e.g. \cite{2008ApJ...677L..77M}).  
As the deviation between N-body simulations and the theoretical 
expectation is stronger for lower mass halos, such an extension of the
high-peak formulation could be seen as a possible resolution.  
We also show that to leading order in $\fnl,\:\gnl,\dots$, the
thresholding approach is equivalent to a local bias expansion.  

In the first PBS approach 
\cite{2008PhRvD..77l3514D,2008JCAP...08..031S,fsmk}, we decompose the 
non-Gaussian perturbations into parts that are linear, quadratic, and cubic 
in Gaussian fields.  We separate long- from short-wavelength perturbations
(the two are uncorrelated for Gaussian fields, but correlated in the
non-Gaussian case), and calculate the bias as the response of the
halo number density to a long-wavelength density perturbation.  
This approach is conceptually simple and offers a clear physical picture 
of the impact of primordial non-Gaussianity on the clustering of biased 
tracers, by isolating the effect of the mode-coupling induced by
non-Gaussianity.  For example, for a generic primordial bispectrum, the 
variance of the small-scale density field is locally rescaled by 
long-wavelength potential fluctuations. Depending on the exact shape of the 
bispectrum,  this rescaling can be scale-independent 
(local model of non-Gaussianity), which then leads to a scale-dependent bias
as in \refeq{one};  or scale-dependent (e.g. orthogonal and equilateral 
models), generally softening the $1/k^2$-dependence.  We will see that
for cubic-order non-Gaussianity, long wavelength perturbations 
not only rescale the local variance of the density field, but also induce 
a local skewness.  Since the abundance of halos also depends on the 
skewness of the small-scale density field (a fact exploited when searching
for non-Gaussianity using the mass function of e.g. galaxy clusters),
this effect contributes to the non-Gaussian halo bias.  
This first PBS approach has the advantage that it can be generically 
applied to any prescription for the average halo abundance (mass function).  
On the other hand, it assumes a clear separation between 
long- and short- wavelength modes, which breaks down when measuring the
clustering on sufficiently small scales.  

Our second PBS approach is inspired by a calculation of the scale-dependent
bias factors in the Gaussian peaks model \cite{2010PhRvD..82j3529D}
(e.g., the first order bias is $\bias{I}(k) = b_{10} + b_{01} k^2$). 
In this approach, we apply the peak-background split directly 
to the non-Gaussian density field.  This is done by calculating the
non-Gaussian conditional mass function using an Edgeworth expansion of 
its Gaussian counterpart.  The halo density contrast is then obtained
by taking the ratio of the unconditional to conditional mass function, 
and expanding with respect to the large-scale density contrast.  This
allows us to determine the linear bias as the lowest-order coefficient
in this series.  This approach can in principle be applied to any
excursion-set mass function.  As a first step, we will here formulate it 
under the assumption that the Press-Schechter multiplicity function 
describes halo abundances. On the other hand, this approach does not
rely on a separation of scales. Thus, the two PBS approaches presented 
here make complementary assumptions.  

The paper is organized as follows: 
we begin by reviewing the models of primordial non-Gaussianity considered
here and spelling out our notation in \refsec{prelim}. 
We discuss the thresholding approach to non-Gaussian bias and point out
its limitations in \refsec{MLB}. \refsec{GPBS} introduces the first PBS 
approach based on a separation of scales, while \refsec{CPBS} 
presents the second PBS approach based on conditional mass functions.  
\refsec{PBS-thr} presents a comparison of the PBS and thresholding
approaches.  The comparison of our predictions with the results of
N-body simulations is the subject of a companion Letter
\cite{2011PhRvD..84f1301D}.  We conclude in \refsec{conclusion}. 


\section{Preliminaries}
\label{sec:prelim}

\subsection{Four types of primordial non-Gaussianity}

Throughout the paper, we will apply our results to the following four 
models of primordial non-Gaussianity (NG).  We parameterize primordial NG via 
the $N$-point functions  ($N>3$) of the Bardeen potential $\Phi(\bm{x})$,
a relativistic generalization of the Newtonian gravitational potential,
in the matter-dominated era.  
Note that $\Phi(\bm{x})$ has the opposite sign relative to the usual 
Newtonian gravitational potential.  Since our goal is to compare analytic
predictions with the outcome of N-body simulations, our set of models 
includes all the templates for which simulations have been performed. Our 
main theoretical results, however, will always be given in terms
of general $N$-point functions and can be straightforwardly applied
to any given model of non-Gaussianity.  

\subsubsection{Local non-Gaussianity}
\label{sec:localNG}

In local primordial NG, the non-Gaussian field $\Phi$ is defined by 
a local Taylor expansion around a Gaussian random field $\phi$ as 
\cite{1990PhRvD..42.3936S,1994ApJ...430..447G,2000MNRAS.313..141V,2001PhRvD..63f3002K}
\be
\label{eq:localNG}
\Phi(\vx)=\phi(\vx)+\fnl\phi^2(\vx)+\gnl\phi^3(\vx)\;.
\ee
Here, $\fnl$ and $\gnl$
are dimensionless, phenomenological parameters which we seek to constrain
using cosmic microwave background (CMB) or large-scale structure (LSS) 
observations. 
This type of non-Gaussianity is typically produced in inflationary models
with more than one scalar field.
Since the primeval curvature perturbations are of magnitude 
${\cal O}(10^{-5})$, the cubic order correction is negligibly 
small compared to the quadratic one when ${\cal O}(\fnl)\sim{\cal O}(\gnl)$. 
However, this condition is not satisfied by some multi-field models such as 
the curvaton scenario, in which a large $\gnl$ and a small $\fnl$ can be 
simultaneously produced \cite{2006PhRvD..74j3003S,2008JCAP...09..012E,
2008JCAP...09..025H,2009JCAP...06..035H,2009JCAP...08..016B}. At leading 
order, the quadratic term generates a 3-point function or bispectrum, 
\be
\label{eq:Bloc}
\xi_\Phi^{(3)}\!(\vk_1,\vk_2,\vk_3)=
2\fnl\Bigl[P_\phi(k_1)P_\phi(k_2)+\mbox{(2 cyc.)}\Bigr]\;,
\ee
where (cyc.) denotes cyclic permutations of the indices,
$P_\phi(k)\propto k^{n_s-4}$ is the power spectrum of the Gaussian field 
$\phi(\vx)$, and $n_s$ is its logarithmic slope. On the other hand, the 
cubic-order terms generate a 4-point function or trispectrum,
\be
\label{eq:Tloc}
\xi_\Phi^{(4)}\!(\vk_1,\vk_2,\vk_3,\vk_4)=
6\gnl\Bigl[P_\phi(k_1) P_\phi(k_2) P_\phi(k_3)+\mbox{(3 cyc.)}\Bigr]\;.
\ee
Both bispectrum and trispectrum are peaked on squeezed triangle or
quadrilateral configurations, i.e. configurations where one side
$|\vk_i|$ is much shorter than the other sides.

\subsubsection{Scale-dependent $\fnl$}

Next, we will 
consider a model in which the quadratic coupling dominates but $\fnl$ 
is $k$-dependent. The primordial bispectrum takes the form 
\citep{2010arXiv1010.3722S}
\be
\label{eq:Block}
\xi_\Phi^{(3)}\!(\vk_1,\vk_2,\vk_3)=\fnl(k_1) P_\phi(k_2) P_\phi(k_3)
+\mbox{(5 perm.)}\qquad \mbox{with}\qquad
\fnl(k)=\fnl(k_p)\left(\frac{k}{k_p}\right)^{n_f}\;,
\ee
where $k_p$ is some arbitrary fixed scale, and $n_f$ is a 
spectral index. 

\subsubsection{Folded and orthogonal non-Gaussianity}

As a third template, we will consider the folded or flattened shape, for 
which the primordial bispectrum reads \cite{2009JCAP...05..018M}
\begin{align}
\nonumber
\xi_\Phi^{(3)}\!(\vk_1,\vk_2,\vk_3) &= 
6\fnl\Bigl[\bigl(P_\phi(k_1) P_\phi(k_2)+\mbox{(2 cyc.)}\bigr)
+3\bigl(P_\phi(k_1) P_\phi(k_2) P_\phi(k_3)\bigr)^{2/3} \\
&\qquad\qquad -\bigl(P_\phi(k_1)^{1/3} P_\phi(k_2)^{2/3} P_\phi(k_3)
+\mbox{(5 perm.)}\bigr)\Bigr] \;.
\label{eq:Bfol}
\end{align}
The folded shape approximates the non-Gaussianity due to modification of the 
initial Bunch-Davies vacuum in canonical single field inflation (the actual 
3-point function is not factorizable).  This template induces 
a scale-dependent bias on large scales with somewhat weaker $k$-dependence
than the local model. \cite{2009ApJ...706L..91V,fsmk}. The 
orthogonal template introduced by \cite{2010JCAP...01..028S},
\begin{align}
\nonumber
\xi_\Phi^{(3)}\!(\vk_1,\vk_2,\vk_3) &= 
6\fnl\Bigl[-3\bigl(P_\phi(k_1) P_\phi(k_2)+\mbox{(cyc.)}\bigr)
-8\bigl(P_\phi(k_1) P_\phi(k_2) P_\phi(k_3)\bigr)^{2/3} \\
&\qquad\qquad +3\bigl(P_\phi(k_1)^{1/3} P_\phi(k_2)^{2/3} P_\phi(k_3)
+\mbox{(5 perm.)}\bigr)\Bigr] \;,
\label{eq:Bort}
\end{align}
gives rise to a similar non-Gaussian halo bias \cite{fsmk}, but roughly 
twice as large in magnitude and opposite in sign (for fixed $\fnl$) 
~\footnote{Even though this shape does not correspond to any physical
mechanism generating primordial NG, we will consider it since it has 
been simulated \cite{2011arXiv1102.3229W}.}

\subsubsection{Equilateral non-Gaussianity}

Finally, the equilateral type of non-Gaussianity, which arises in inflationary
models with higher-derivative operators such as the DBI model, is well
described by the factorizable form \cite{2006JCAP...05..004C}
\begin{align}
\label{eq:Beq}
\xi_\Phi^{(3)}(\vk_1,\vk_2,\vk_3) &=6\fnl
\Bigl[-\bigl(P_\phi(k_1)P_\phi(k_2)+\mbox{(cyc.)}\bigr)
-2\bigl(P_\phi(k_1)P_\phi(k_2)P_\phi(k_3)\bigr)^{2/3}\Bigr.  \\
& \qquad\qquad +  \Bigl. \bigl(P_\phi(k_1)^{1/3}P_\phi(k_2)^{2/3} P_\phi(k_3)
+\mbox{(5 perm.)}\bigr)\Bigr]\nonumber \;.
\end{align}
It can easily be verified that the signal is largest in the equilateral
configurations $k_1\approx k_2\approx k_3$, and suppressed in the
squeezed limit $k_3\ll k_1\approx k_2$.

\subsection{From primordial perturbations to galaxies}

In standard CDM cosmologies, galaxies form inside dark matter halos and 
this introduces a bias between the mass and the galaxy distributions
\cite{Kaiser1984}. In what follows, we shall adopt a Lagrangian picture. 
Namely, we express the clustering of biased tracers, such as 
dark matter halos of mass $M$ collapsing at redshift $z$, in terms of the
statistics of the \emph{initial} density perturbation $\delta_R(\vk,z)$ 
smoothed on a scale $R$ and linearly evolved to redshift $z$, where $R$ is
related to $M$ via $M = (4\pi/3)\rhob R^3$.  More precisely, $\d$ is
the fractional density perturbation in synchronous gauge.  Thus, the Poisson 
equation provides a relationship between $\delta_R(\vk,z)$ and the Bardeen 
potential $\Phi(\vx)$ via 
\be
\label{eq:poisson}
\delta_R(\vk,z)={\cal M}_R(k,z)\Phi(\vk)\;,
\ee
where
\be
\label{eq:transfer}
{\cal M}_R(k,z)= \Mm(k,z) W_R(k) = \frac{2}{3}
\frac{k^2 T(k) g(z)}{\Omega_{m} H_0^2 (1+z)}W_R(k)\;.
\ee
Here, $T(k)$ is the matter transfer function normalized to unity as $k\to 0$, 
$g(z)$ is the linear growth rate of the gravitational potential normalized 
to unity during the matter dominated epoch, and $W_R(k)$ is a (spherically 
symmetric) window function with characteristic radius $R$. We will assume a
spherical top-hat filter throughout. Note 
also that the matter power spectrum at redshift $z$ is related to the 
primordial curvature power spectrum through $P_m(k,z) = \Mm^2(k,z) P_\phi(k)$.

Regardless of the initial conditions, we shall denote the Lagrangian bias 
factors of dark matter halos by $\bo$, $\bt$, ..., while Eulerian bias 
parameters are denoted as ${\rm b}_{\rm I}^{\rm E}$, etc. Note that these 
bias parameters are generally scale-dependent. The notation $b_1$, $b_2$, 
$b_1^{\rm E}$, etc. will exclusively designate the {\it Gaussian, 
scale-independent} peak-background split biases.  In the next Section, we 
will also use the notation $c_1$, $c_2,\dots$ for the mass-weighted, 
cumulative Gaussian bias parameters which appear in the thresholding 
approach.  

We will describe the abundance of halos through their mass function
$n_h\equiv dn/dM$ which we will assume to be of the universal form, 
i.e.
\be
\label{eq:umassfn}
\nh=\frac{\rhob}{M^2} f(\nu) 
\left|\frac{\partial\ln\s_{0M}}{\partial\ln M}\right|\;,
\ee
where $f(\nu)$ is the multiplicity function and $\s_{0M}$ is the
RMS density fluctuation on scale $M$.  

Unless otherwise specified, we shall adopt in all illustrations a flat 
$\Lambda$CDM cosmology with $\Om = 0.279$, $h=0.7$, and an adiabatic 
initial perturbations with spectral index $n_s=0.96$ and 
amplitude $A_s=7.96\times 10^{-10}$ at the pivot point $k_0=0.02$~Mpc$^{-1}$ 
(corresponding to a normalization $\s_8\approx 0.81$). These values are 
consistent with the latest CMB constraints from
WMAP7 \cite{2011ApJS..192...18K}.


\section{Statistics of thresholded regions}
\label{sec:MLB}

In this Section, we shall present the derivation of the scale-dependent 
non-Gaussian bias using the statistics of regions above threshold 
\citep{1986ApJ...310L..21M,GrinsteinWise}, without invoking the high 
threshold (high peak) approximation.  Several concepts and results 
introduced in this Section will be employed later in the paper.

\subsection{Probability densities}

In the Press-Schechter approach \cite{1974ApJ...187..425P}, virialized 
objects are identified with high-density regions in the linear density 
field. The two-point correlation function of thresholded regions,
$\xnu(r)$, can be calculated once the probability $P_1$ of finding a region 
whose  overdensity is above the threshold 
$\delta_c\approx 1.69$ 
\cite{1972ApJ...176....1G}, and the probability $P_2$ of finding two such 
regions separated by a distance $r\equiv |\vx_2-\vx_1|$, are known.  It
is convenient to express the results in terms of the significance
(peak height) $\nu \equiv \d_c/\s_{0s}$, where $\s_{0s}$ is the r.m.s. 
variance of the density field smoothed on scale $R_s$.  The correlation
function is then given by \cite{Kaiser1984}:
\be
\xnu(r) = \frac{P_2(>\nu,r)}{\bigl[P_1(>\nu)\bigr]^2} -1.
\ee
$\xnu(r)$ is commonly interpreted as describing the 
2-point correlation of halos above mass $M$ corresponding to the 
smoothing length $R_s$.  For any 
non-Gaussian initial density field, $P_1$ and $P_2$ can be expressed in 
terms of the $N$-point connected correlation functions as follows
~\footnote{A complete derivation can be found in Appendix K of 
\cite{jeong:2010}.}:
\begin{align}
P_1(>\nu)
=&
\frac{1}{\sqrt{2\pi}}
\int_\nu^\infty \!\! dy\,
\exp\left[
\sum_{N=3}^\infty
(-1)^N\frac{w_s^{(N,0)}}{N!}\frac{d^N}{dy^N}
\right]
e^{-y^2/2}
\label{eq:MLB_P1}
\\
P_2(>\nu,r)
=&
\frac{1}{2\pi}
\int_{\nu}^\infty\!\! dy_1
\int_{\nu}^\infty\!\! dy_2\,
\exp\left[
\sum_{N=2}^\infty
\sum_{m=0}^N (-1)^N\frac{w_s^{(N,m)}(r)}{m!(N-m)!}
\frac{\partial^N}{\partial y_1^m\partial y_2^{N-m}}
\right]
e^{-\frac{1}{2}(y_1^2+y_2^2)}
\label{eq:MLB_P2}
\end{align}
For shorthand convenience, we will hereafter omit the explicit 
$z$-dependence of $\d_s(\vx)\equiv\d_{R_s}(\vx)$ and 
$\Mm_s(k)\equiv\Mm_{R_s}(k)$.  We have also defined
\be
w_s^{(N,m)}(r)\equiv
\left\{
\begin{array}{ll}
w_s^{(2,m)} = \xi_s^{(2,m)}(r)/\sigma_{0s}^2 &(m=1)
\\
w_s^{(2,m)} = 0  &(m=0\mathrm{~or~}2)
\\
w_s^{(N,m)} = \xi_s^{(N,m)}(r)/\sigma_{0s}^N &(N>2)
\end{array}
\right.,
\ee
where
\be
\xi_s^{(N,m)}(r)
\equiv \Big\langle
\underbrace{
\delta_s(\bm{x}_1)\cdots\delta_s(\bm{x}_1)
}_{m~\textrm{times}}
\underbrace{
\delta_s(\bm{x}_2)\cdots\delta_s(\bm{x}_2)
}_{N-m~\textrm{times}}\Big\rangle_c
\ee
is the $N$-point connected correlation function evaluated at two different 
locations $\bm{x}_1$ and $\bm{x}_2$. Note that the correlation 
$w_s^{(N,0)} = w_s^{(N,N)}$ is evaluated at zero lag, and that the 
probability densities $P_1$ and $P_2$ depend explicitly on the smoothing 
scale $R_s$ through the functions $w_s^{(N,m)}$ and the peak height 
$\nu\equiv\delta_c/\sigma_{0s}$.

\subsection{Bias parameters for a Gaussian density field}

It is instructive to first perform the calculation for Gaussian 
initial conditions.  Later, we shall use the Gaussian bias derived in this
Section to identify the coefficients of the non-Gaussian scale dependent bias.

\subsubsection{Gaussian bias factors from a peak-background split}

When the underlying smoothed density field obeys the Gaussian statistics, 
the probability $P_1$ of exceeding the threshold $\nu$ is given by
\begin{align}
\label{eq:def_P1G}
P_1(>\nu) =& 
\frac{1}{\sqrt{2\pi}}
\int_\nu^\infty\!\! dx\, e^{-x^2/2} 
=
\frac{1}{2}\,\erfc\left(\frac{\nu}{\sqrt{2}}\right).
\end{align}
In the peak-background split approach, one considers the effect of adding a 
long-wavelength (background) perturbation $\delta_l$ of 
characteristic wavelength $R_l\gg R_s$
to the small scale density field (peak) $\delta_s$.  
Assuming that $\delta_l$ is independent of $\delta_s$.
it is clear that adding $\d_l$ is equivalent to reducing the threshold 
$\nu \rightarrow (\d_c-\d_l)/\s_{0s}$; thus, $P_1(>\nu,\delta_l)$, the 
probability $P_1$ in the large-scale overdensity $\delta_l$ is given by
\begin{equation}
P_1(>\nu,\delta_l) 
=
P_1
\left(>\nu-\frac{\delta_l}{\sigma_{0s}}\right).
\end{equation}
We define the peak-background split 
bias factors $c_N$ as the fractional change of $P_1$ with $\d_l$ via
\begin{equation}
\label{eq:def_cN}
c_N \equiv \frac{1}{P_1(>\nu)} \frac{d^N\, P_1(>\nu, \d_l)}{d\,\d_l^N}\;,
\end{equation}
so that
\begin{align}
\label{eq:c_N}
c_N(\nu) 
=&
\left(-\frac{1}{\sigma_{0s}}\right)^N
\frac{1}{P_1(>\nu)}
\frac{d^N\!\bigl[P_1(>\nu)\bigr]}{d\nu^N}
=
\sqrt{\frac{2}{\pi}}
\left[
\erfc\left(\frac{\nu}{\sqrt{2}}\right)\right]^{-1}
\frac{e^{-\nu^2/2}}{\sigma_{0s}^N}
H_{N-1}\left(\nu\right).
\end{align}
Here, $H_N$ is the Hermite polynomial defined by
\be
H_N(x) \equiv (-1)^N e^{x^2/2} \frac{d^N}{dx^N}
\left(
e^{-x^2/2}
\right).
\ee
Note that we adopt the so-called \textit{probabilists'} convention for the
Hermite polynomials.  It is related to the so-called \textit{physicists'} 
convention by
\be
H_N^\mathrm{phys}(x) = 2^{N/2} H_N\bigl(\sqrt{2}x\bigr)\;.
\ee
Explicit expressions for the first five Hermite polynomials are
\be
H_0(x) = 1,\qquad
H_1(x) = x,\qquad
H_2(x) = x^2-1, \qquad
H_3(x) = x^3-3x,\qquad
H_4(x) = x^4-6x^2+3 \;.
\ee
Since $H_N\to\nu^N$ for large $\nu$, we see that in the high-peak limit 
($\nu\gg1$), 
\be
c_N \approx \nu H_{N-1}(\nu)/\sigma_{0s}^N \approx \nu^N/\sigma_{0s}^N;.
\label{eq:cNhp}
\ee

The one-point probability \refeq{def_P1G} and bias parameters 
\refeq{c_N} are \emph{cumulative}. They describe the number density and 
bias of all peaks above the threshold $\nu$ at fixed smoothing scale 
$R_s$.  
To relate these quantities to the mass function and bias of dark matter
halos, we follow Press \& Schechter \cite{1974ApJ...187..425P} and 
interpret $P_1$ as the fraction of the Lagrangian volume occupied by halos 
of mass exceeding $M$.  Therefore, the halo number density follows upon
dividing the derivative of $P_1$ w.r.t. mass by $M/\bar{\rho}$,
\be
\bar{n}_h(M) = 
-2 \frac{\bar\rho}{M} \frac{d}{dM} 
P_1 \left(>\nu\right)
= 2\frac{\bar\rho}{M^2} \frac{\nu e^{-\nu^2/2}}{\sqrt{2\pi}} 
\biggl\lvert\frac{d \ln \sigma_{0s}}{d\ln M}\biggr\lvert\;,
\label{eq:nPS}
\ee
where the factor of 2 is introduced to account for the fact that regions
with $\d<\d_c$ may be embedded in regions with $\d>\d_c$ on scale $>R_s$
(clouds-in-clouds).  Thus, \refeq{nPS} is of the form \refeq{umassfn} with
$f(\nu) = \sqrt{2/\pi} \nu\exp(-\nu^2/2)$.  Conversely, integrating
\refeq{nPS} yields
\be
\label{eq:P1_cul_MnM}
P_1(>\nu) 
=
\frac{1}{2\bar\rho}
\int_M^\infty dM' M' \bar{n}_h(M')\;.
\ee
Inserting this into \refeq{def_cN}, we find that the $c_N$ are 
mass-weighted cumulative bias factors,
\be
c_N = 
\left[
\int_M^\infty dM' M' \bar{n}_h(M')
\right]^{-1}
\int_M^\infty dM' M' \bar{n}_h(M') b_N(M')\;,
\ee 
where 
\be
b_N(M) = \frac{1}{\nu_M}\frac{H_{N+1}(\nu_M)}{\sigma_{0M}^N}
\label{eq:bNdef}
\ee
are the peak-background split biases derived from the Press-Schechter
mass function. Here, $\nu_M$ and $\sigma_{0M}$ denote the significance and
 r.m.s. density fluctuation on the mass scale $M$.  
It is only in the high-peak limit ($\nu\gg1$) that the mass-weighted 
cumulative bias $c_N$ and the bias $b_N(M)$ asymptote to the same values
[\refeq{cNhp}].  

So far, we have not yet specified any prescription for how to go 
from the bias parameter $c_N$ to the clustering of tracers.  
This will be elucidated in the next Section, where we calculate
the correlation function of thresholded regions directly.  

\subsubsection{Gaussian bias factors from the correlation of thresholded regions}
\label{sec:xinu}

In this Section, we present the calculation of the 
two-point correlation function $\xnu(r)$ of thresholded regions assuming 
Gaussian initial conditions, and show that the cumulative mass-weighted 
biases $c_N$ obtained with the peak-background split coincide with the bias 
parameters arising in $\xnu(r)$.
 
Observing that, for Gaussian initial conditions, all the connected 
correlation functions $\xi_s^{(N,m)}$ with $N>2$ vanish, we can express
$\xnu(r)$ as 
\begin{align}
\nonumber
\xnu(r) 
\equiv\frac{P_2(>\nu,r)}{[P_1(>\nu)]^2} -1
=
\frac{2}{\pi}\left[\erfc\left(\frac{\nu}{\sqrt{2}}\right)\right]^{-2}
\int_\nu^\infty\!\! dy_1
\int_\nu^\infty\!\! dy_2\,
\exp\left[
\frac{\xi_s(r)}{\sigma_{0s}^2}
\frac{\partial^2}{\partial y_1\partial y_2}
\right]
e^{-\frac{1}{2}\left(y_1^2+y_2^2\right)}-1 \;.
\end{align}
Here, $\xi_s(r)\equiv\xi_s^{(2,1)}$ is the 2-point density correlation 
smoothed on scale $R_s$.  On employing the definition of $H_N(x)$, we can 
further simplify the double integration as 
\begin{align}
&\int_\nu^\infty\!\! dy_1
\int_\nu^\infty\!\! dy_2\,
\exp\left[
\frac{\xi_s(r)}{\sigma_{0s}^2}
\frac{\partial^2}{\partial y_1\partial y_2}
\right]
e^{-\frac{1}{2}\left(y_1^2+y_2^2\right)}
=\frac{\pi}{2}\left[\erfc\left(\frac{\nu}{\sqrt{2}}\right)\right]^2
+
\sum_{N=1}^\infty
\frac{\left[\xi_s(r)\right]^N}{N!\sigma_{0s}^{2N}}
\left[H_{N-1}(\nu)\right]^2
e^{-\nu^2/2}\;.
\end{align}
Therefore, we find that the 2-point correlation function of thresholded 
regions is given by \cite{JensenSzalay86}
\be
\xnu(r) 
=
\frac{2}{\pi}\left[\erfc\left(\frac{\nu}{\sqrt{2}}\right)\right]^{-2}
\sum_{N=1}^\infty
\frac{\bigl[\xi_s(r)\bigr]^N}{N!\sigma_{0s}^{2N}}
\bigl[H_{N-1}(\nu)\bigr]^2
e^{-\nu^2/2}\;.
\ee
Next, on substituting the expression of the cumulative peak-background split 
bias factors \refeq{c_N}, we can recast the peak correlation function into 
the series
\be
\xnu(r)
=
\sum_{N=1}^{\infty}
\frac{c_N^2}{N!}\bigl[\xi_s(r)\bigr]^N\;.
\label{eq:xinu}
\ee
If we compare the expression for $\xnu(r)$ to that obtained from a
local bias expansion \cite{1993ApJ...413..447F} of the density $\d_{>\nu}$ 
of regions above threshold,
\be
\d_{>\nu}(\bm{x}) = \sum_{N=1}^\infty \frac{\tilde{c}_N}{N!}\,
\bigl[\d_s(\bm{x})\bigr]^N,
\ee
we see that the coefficient $\tilde{c}_N$ is different from the $c_N$ 
appearing in the correlation function:  when
calculating $\xi_{>\nu} = \<\d_{>\nu}(\vx_1)\d_{\nu}(>\vx_2)\>$, the 
coefficient of $\left[\xi_s(r)\right]^N$ includes not only $\tilde{c}_N^2$, 
but also terms such as $\tilde{c}_{N}\tilde{c}_{N+2m}\sigma_{0s}^{2m}$ for 
all positive integers $m \leq N/2$.  
This clearly shows that the bias parameters $c_N$ from the peak-background 
split are to be seen as ``renormalized'' bias parameters 
\cite{2011PhRvD..83h3518M} which take all the 
higher order moments into account, and thus truly are the coefficients of 
the observed correlation function of (in this case) thresholded regions.

\subsection{Two-point correlation function of thresholded regions with non-Gaussianity}

In the presence of primordial non-Gaussianity, all the correlation functions
$\xi_s^{(N,m)}$ are in principle necessary to determine $P_1(>\nu)$ and
$P_2(>\nu,r)$.  Here, we will restrict ourselves to the leading order
corrections linear in the correlations $\xi_s^{(N,m)}$.  We derive
a general expression for the scale-dependent non-Gaussian bias induced 
by a primordial $N$-point function $\xi_\Phi^{(N)}$.  

\subsubsection{Relation to local deterministic bias}

First, we show that the leading order contribution to the 
two point correlation function of thresholded regions, which includes terms 
linear in the connected correlations functions $\xi_s^{(N,m)}$ only, is
consistent with the result from a local deterministic bias ansatz.  
Linearizing the exponential factors in Eqs. (\ref{eq:MLB_P1}) 
and (\ref{eq:MLB_P2}), we obtain
\begin{align}
P_1(>\nu)
\approx &
\frac{1}{2}\,\erfc\left(\frac{\nu}{\sqrt{2}}\right)
+
\sum_{N=3}^\infty
\frac{1}{\sqrt{2\pi}}
\frac{w_s^{(N,0)}}{N!}
H_{N-1}(\nu)
e^{-\nu^2/2}
\label{eq:MLB_P1_linear}
\\
P_2(>\nu,r)
\approx &
\left[\frac{1}{2}\,\erfc\left(\frac{\nu}{\sqrt{2}}\right)\right]^2
+
\sqrt{\frac{1}{2\pi}}\erfc\left(\frac{\nu}{\sqrt{2}}\right)
\sum_{N=3}^\infty
\frac{w_s^{(N,0)}}{N!}
H_{N-1}(\nu)
e^{-\nu^2/2} 
\nonumber
\\
&\quad +
\frac{1}{2\pi}
\sum_{N=2}^\infty
\sum_{m=1}^{N-1} 
\frac{w_s^{(N,m)}}{m!(N-m)!}
H_{m-1}(\nu)H_{N-m-1}(\nu)
e^{-\nu^2} \;,
\label{eq:MLB_P2_linear}
\end{align}
where we have neglected terms beyond linear order.  Thus, the
two-point correlation function of thresholded regions reads
\begin{align}
\nonumber
\xnu(r) 
&= 
\frac{2}{\pi}
\left[\erfc\left(\frac{\nu}{\sqrt{2}}\right)\right]^{-2}
\sum_{N=2}^\infty
\sum_{m=1}^{N-1} 
\frac{w_s^{(N,m)}}{m!(N-m)!}
H_{m-1}(\nu)H_{N-m-1}(\nu)
e^{-\nu^2}
\\
&=
\sum_{N=2}^\infty
\sum_{m=1}^{N-1} 
\frac{c_m c_{N-m}}{m!(N-m)!}\,
\xi_s^{(N,m)}(r)\;.
\label{eq:MLB_xip_nG}
\end{align}
As can easily be seen, a local deterministic mapping
\be
\label{eq:localbias}
\d_{>\nu}(\bm{x})
=
\sum_{N=0}^\infty \frac{c_N}{N!}\bigl[\delta_s(\bm{x})\bigr]^N,
\ee
yields the same result at leading order (the renormalization of the bias 
parameters $c_N$ discussed in the previous Section for the Gaussian case 
will apply at second and higher order).  
This shows that, at first order in $\xi_s^{(N,m)}$, the correlation of 
thresholded regions with primordial non-Gaussianity is equivalent to a 
local deterministic bias relation.  
Note that, for non-Gaussian initial conditions, an effective first-order 
bias defined through $c_{1,\rm eff}\equiv\sqrt{\xnu(r)/\xi_s(r)}$ is 
generally scale-dependent.

\subsubsection{Power spectrum of thresholded regions}

We now Fourier-transform \refeq{MLB_xip_nG}, and investigate the
separate terms.  For simplicity and without loss of generality, we will 
assume that a single non-Gaussian $N$-point function ($N>3$) dominates.  
We then have
\be
P_{>\nu}(k) = c_1^2 P_s(k) + \sum_{m=1}^{N-1} \frac{c_m c_{N-m}}{m! (N-m)!}
\tilde{\xi}_s^{(N,m)}(k)\;.\label{eq:Pknu}  
\ee
Here, $P_s(k) = W_{R_s}^2(k) P(k)$ is the matter power spectrum smoothed
on scale $R_s$.  Let us consider the term $m=1$ first.  It is in fact
identical to the term  $m=N-1$.  We have
\ba
\tilde{\xi}_s^{(N,1)}(k) =\:& \prod_{i=1}^{N-1}\left(\int\!\!\frac{d^3k_i}{(2\pi)^3}\right)
\xi_s^{(N)}(\vk,\vk_1,\dots,\vk_{N-1})\; (2\pi)^3 \d_D(\vk+\vk_1+\dots+\vk_{N-1})\vs
=\:& \Mm_s(k) \prod_{i=1}^{N-2}\left(\int\!\!\frac{d^3k_i}{(2\pi)^3} \Mm_s(k_i)\right)
\Mm_s(q)\: \xi_\Phi^{(N)}(\vk,\vk_1,\dots,\vk_{N-2},\v{q};X)\;.
\label{eq:xiN1}
\ea
Here, $\v{q} = - \vk_1 -\dots-\vk_{N-2}-\vk$, and $X$ is a set of
variables characterizing the primordial $N$-point function such as 
$\fnl,\gnl,n_f$ depending on the details of the model of non-Gaussianity.  
In the second line, we have used the fact that the matter $N$-point function
is related to the $N$-point function of the potential $\Phi$ through
\be
\xi_s^{(N)}(\bm{k}_1,\dots,\bm{k}_N) =
\left(\prod_{i=1}^N\Mm_s(k_i)\right)  
\xi_\Phi^{(N)}(\bm{k}_1,\dots,\bm{k}_N;X)\;.
\ee
Note that the scaling of $\tilde{\xi}_s^{(N,1)}$ in the large-scale limit
($k\to 0$) depends on the scaling of $\xi_\Phi^{(N)}$ in the 
\emph{squeezed limit}, where one argument ($k$) is much smaller than
the others ($k_1,\dots,k_{N-2},q$).  

Next, consider the term with $m=2$ (again, it is equal to the term $m=N-2$).  
A similar calculation leads to
\ba
\tilde{\xi}_s^{(N,2)}(k) =\:& \prod_{i=1}^{N-1}\left(\int\!\!\frac{d^3k_i}{(2\pi)^3}\right)
\xi_s^{(N)}(\vk-\vk_1,\vk_1,\dots,\vk_{N-1})\; (2\pi)^3 \d_D(\vk+\vk_2+\dots+\vk_{N-1})\vs
=\:& \prod_{i=1}^{N-2}\left(\int\!\!\frac{d^3k_i}{(2\pi)^3} \Mm_s(k_i)\right)
\Mm_s(|\vk-\vk_1|) \: \Mm_s(q)\: \xi_\Phi^{(N)}(\vk-\vk_1,\vk_1,\dots,\vk_{N-2},\v{q};X)\;,
\label{eq:xiN2}
\ea
where now $\v{q} = -\vk-\vk_2-\dots-\vk_{N-2}$.  In the large-scale 
(small-$k$) limit, $|\vk-\vk_1| \gg k$, so that $\tilde{\xi}_s^{(N,2)}$ 
approaches a constant.  One can easily verify that this also holds
for all $m\geq 3$ terms.  On large scales, these terms thus all
add white-noise contributions to the power spectrum of thresholded
regions, and only the terms with $m=1,\:N-1$ contribute to the 
scale-dependent bias.  Note, however, that the white-noise corrections 
which appear for non-Gaussianity of order $N\geq 4$ can be quite 
significant \cite{2010PhRvD..81b3006D}.  

A general feature of the non-Gaussian power spectrum of tracers in the
thresholding approach is now clear:  the presence of a primordial $N$-point
function generates a dependence of $P_{>\nu}(k)$ on the bias $c_{N-1}$
through $\tilde{\xi}_s^{(N,1)}(k)$, which depends on the scaling
of $\xi_\Phi^{(N)}$ in the squeezed limit.    
The former has also been pointed out by \citep{2008PhRvD..78l3534T,
2009PhRvD..80l3002S,2009ApJ...703.1230J}, who studied the non-Gaussian 
bias in the local, constant-$\fnl$ model.  As shown here, this conclusion 
also holds if we were to consider any local bias mapping of the form 
\refeq{localbias} (at leading order in the non-Gaussian $N$-point function).  
We can then rewrite \refeq{Pknu} as
\ba
P_{>\nu}(k) =\:& c_1^2 P_s(k) + 2\frac{c_1 c_{N-1}}{(N-1)!}
\tilde{\xi}_s^{(N,1)}(k)\vs
=\:& \left[ c_1^2 + 2\frac{4}{(N-1)!} c_1 c_{N-1} \s_{0s}^2 \Mm_s^{-1}(k) 
{\cal F}_s^{(N)}(k, X)\right] P_s(k)\;.\label{eq:pnu_final}
\ea
The factor of 2 comes from the sum of the $m=1,\:N-1$ terms, and we have 
introduced the shape factor
\ba
{\cal F}_R^{(N)}(k,X) \equiv\:& \frac{\Mm^{-1}_R(k)}{4\s_{0R}^2 P_\phi(k)} 
\tilde{\xi}_R^{(N,1)}(k)\vs
=\:& \frac{1}{4\sigma_{0R}^2\,P_\phi(k)}
\biggl\{\prod_{i=1}^{N-2}\int\!\!\frac{d^3k_i}{(2\pi)^3}\Mm_R(k_i)\biggr\}
\Mm_R(q)\: \xi_\Phi^{(N)}(\bm{k}_1,\cdots,\bm{k}_{N-2},\bm{q},k\zvh; X)\;,
\label{eq:fshape}
\ea
where $\zvh$ is some arbitrary unit vector.
Noting that $P_{>\nu} = (c_1^2 + 2 c_1 \D c_1) P_s$ to leading order in the 
non-Gaussian corrections, we can identify the scale-dependent correction
to the linear bias as
\be
\Delta c_1(k)
= 
\frac{4 c_{N-1}}{(N-1)!} 
\s_{0s}^2
\frac{{\cal F}_s^{(N)}(k)}{\Mm_s(k)}\;.
\label{eq:dbk-hp1}
\ee
In the rest of this Section, we derive the non-Gaussian correction to the
clustering of thresholded regions for the four models of primordial NG we 
consider in this paper.  It will prove useful to define general spectral 
moments through 
\be
\label{eq:sigmaAR}
\sigma_{\alpha R}^2\equiv\frac{1}{2\pi^2}\int_0^\infty\!\!dk\,
k^{2(\alpha+1)}\Pp(k)\Mm_R^2(k)\;.
\ee

\subsubsection{Local non-Gaussianity}

For the cubic local model, described by primordial three- and four-point
functions [\refsec{localNG}], \refeq{pnu_final} becomes
\begin{align}
P_{>\nu}(k)
= &
\left[
c_1^2
+
4
c_1 c_2 
\sigma_{0s}^2
\Mm_s^{-1}(k)
{\cal F}_s^{(3)}(k,X)
+
\frac{4c_1c_3}{3}
\sigma_{0s}^2
\Mm_s^{-1}(k)
{\cal F}_s^{(4)}(k,X)
\right]
P_s(k)\;,
\label{eq:pnu_FNR}
\end{align}
where ${\cal F}_s^{(3)}(k,\fnl)$ is precisely equal to the form factor 
introduced by \citep{2008ApJ...677L..77M,2009ApJ...706L..91V}.  
Focusing on the quadratic case first, note that on 
large scales, ${\cal F}_s^{(3)} \simeq \fnl $ and the power 
spectrum for $\gnl=0$ becomes
\be
P_{>\nu}(k) = 
\left[
c_1^2 + 4 \fnl c_1c_2\sigma_{0s}^2 \Mm_s^{-1}(k)
\right] P_s(k).
\ee
From the above equation, it is clear that the scale dependence of the
non-Gaussian bias is $\Delta b(k)\propto\Mm_s^{-1}(k)\propto k^{-2}$.  
For high thresholds $\nu\gg1$ in particular, the Gaussian bias 
parameters $c_N$ approach $\nu^N/\sigma_{0s}^N$ so that we can 
approximate the coefficient of the non-Gaussian correction as 
$c_2\sigma_{0s}^2 \simeq c_1\delta_c$.  Therefore, we recover the 
expression of \cite{2008ApJ...677L..77M},
\be
P_{>\nu}(k) 
\stackrel{\nu\gg1}{=}
b_1^2
\left[
1 + 4 \fnl \frac{\delta_c}{\Mm_s(k)}
\right] P_s(k)\;,
\ee
upon replacing $c_1$ with $b_1$ (i.e., assuming a narrow mass bin).

For the local $\gnl\phi^3$ model, note that 
\begin{align}
P_s(k) &
= \Mm_s^2(k)
\left[
1+6\gnl\sigma_\phi^2
\right]P_\phi(k)
\end{align} 
The matter power spectrum $P_s(k)$ thus contains 
$\sigma_\phi^2\equiv\la\phi^2\ra$, which has a logarithmic divergence for
both large and small scales \cite{2008PhRvD..78l3519M}.  In reality,
the finite survey size and the free-streaming scale of dark matter provide 
low- and high-$k$ cut-offs.  In simulations, 
the finite box size and the resolution provide such 
cutoffs \cite{2010PhRvD..81b3006D}.  

On large scales, the shape factor ${\cal F}_s^{(4)}$ generated by the local 
trispectrum rapidly converges towards $(3/4)\gnl\s_{0s}^2\Sk_{s,\rm loc}$,
where
\be
\label{eq:S3loc}
\Sk_{s,\rm loc} \equiv \frac{6}{\sigma_{0s}^4}
\int\!\!\frac{d^3k_1}{(2\pi)^3} \Mm_s(k_1) P_\phi(k_1)
\int\!\!\frac{d^3k_2}{(2\pi)^3} \Mm_s(k_2) P_\phi(k_2)
\Mm_s(|\bm{k}_1+\bm{k}_2|).
\ee
is the skewness parameter of the density field smoothed on scale $R_s$,
$\<\d_s^3\>/\<\d_s^2\>^2$, in a local $\fnl$ model with $\fnl=1$.  
Therefore, the non-Gaussian contribution to the power spectrum in a pure
$\gnl$ model becomes
\be
\D\pnu(k)\stackrel{k\rightarrow0}{=}
c_1 c_3 \gnl\s_{0s}^4\Sk_{s,\rm loc}\Mm_s(k)\Pp(k)\;.
\ee
Note that the non-Gaussian bias also has a scale-dependence of $k^{-2}$.
For high peaks $\nu\gg 1$, $c_1c_3 \sigma_{0s}^4 = c_1^2\delta_c^2$, and 
we recover Eq.~(21) of \cite{2010PhRvD..81b3006D} upon replacing $c_1$ by 
$b_1$.  In general however, the correct coefficient in the thresholding 
calculation is the third-order bias $c_3$.

\begin{figure*}
\center
\resizebox{0.49\textwidth}{!}{\includegraphics{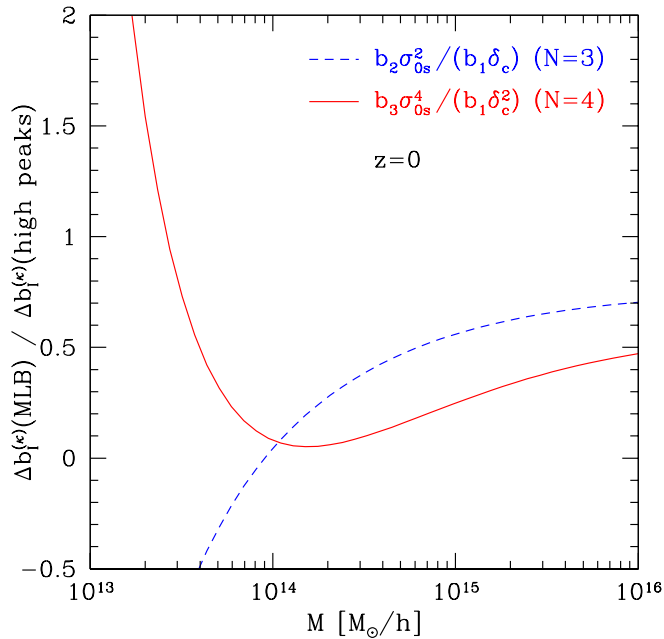}}
\caption{Ratio of the non-Gaussian correction to the linear bias predicted
by the statistics of thresholded regions to that obtained in the high-peak 
limit. For a non-zero primordial bispectrum ($N=3$) and trispectrum ($N=4$), 
this ratio is equal to $b_2\s_{0s}^2/(b_1\d_c)$ and $b_3\s_{0s}^4/(b_1\d_c^2)$,
respectively. Note that it depends on the order $N$ but not on the specific
shape of the primordial correlation function. Results are shown at $z=0$ as 
a function of halo mass $M$. The Gaussian bias parameters $b_N$ are computed 
from a Sheth-Tormen mass function.
\label{fig:emlb}}
\end{figure*}

\subsubsection{Scale-dependent and non-local non-Gaussianity}

For the $k$-dependent local bispectrum Eq.~(\ref{eq:Block}), the 
power spectrum of thresholded regions is
\be
\label{eq:pnu_kfnl}
\pnu(k)=
\left[c_1^2 
+4 c_1 c_2 \sigma_{0s}^2 
\Mm_s^{-1}(k)  {\cal F}_s^{(3)}(k,\fnl,n_f)
\right]
P_s(k)
\ee
where the redshift independent function ${\cal F}_s^{(3)}(k,\fnl,n_f)$ is
computed from Eq.~(\ref{eq:fshape}) on inserting Eq.~(\ref{eq:Block}):
\be
{\cal F}_s^{(3)}(k,\fnl,n_f)
=
\frac{1}{2\sigma_{0s}^2}\frac{\fnl(k_p)}{k_p^{n_f}}
\int\!\!\frac{d^3q}{(2\pi)^3}
\Mm_s(q)
\Mm_s(|\bm{k}-\bm{q}|)
\Pp(q)
\left[
k^{n_f}\frac{\Pp(|\bm{k}-\bm{q}|)}{\Pp(k)}
+2q^{n_f}
\right]
\ee
On large scales, the shape factor converges towards
\be
{\cal F}_s^{(3)}(k,\fnl,n_f)
\stackrel{k\to0}{=}
\frac{1}{\sigma_{0s}^2}\frac{\fnl(k_p)}{k_p^{n_f}}
\int\!\!\frac{d^3q}{(2\pi)^3}
\Mm_s^2(q)
\Pp(q)
q^{n_f}
=
\frac{1}{\sigma_{0s}^2}\frac{\fnl(k_p)}{k_p^{n_f}}
\sigma_{\alpha s}^2,
\ee
where $\sigma_{\alpha s}$ is the spectral moment evaluated for 
$\alpha=n_f/2$. Therefore, the non-Gaussian correction to the peak power 
spectrum becomes
\be
\Delta\pnu(k)\stackrel{k\rightarrow0}{=} 4 c_1 c_2 \fnl(k_p)k_p^{-n_f}
\sigma_{\alpha s}^2 \Mm_s^{-1}(k) P_s(k)\;.
\ee
This result agrees with that of \cite{2010arXiv1010.3722S,fsmk} in the 
high-peak limit only, for which $c_1 c_2\approx (\nu/\sigma_{0s})^3$. 

Finally, for the folded, orthogonal and equilateral bispectrum shapes, the 
power spectrum of thresholded regions is also given by 
Eq.~(\ref{eq:pnu_kfnl}), with ${\cal F}_s^{(3)}$ obtained from an integration 
over \refeq{Bfol} and \refeq{Beq}, respectively \cite{2009ApJ...706L..91V}. 
In the limit $k\ll 1$, we can set $\Mm_s(q)\approx\Mm_s(k_1)$ and, on 
expanding $\Pp(q)$ at second order in $k/k_1$, we arrive at 
\be
\Delta\pnu(k) \stackrel{k\rightarrow0}{=} 
6Ac_1 c_2 \fnl \sigma_{\alpha s}^2 k^{-2\alpha}\Mm_s(k)\Pp(k)\;,
\ee
with $A=1$, $\alpha=(n_s-4)/6\approx -1/2$ (folded shape), $A=-2$, 
$\alpha=(n_s-4)/6\approx -1/2$ (orthogonal shape) and $A=2$, 
$\alpha=(n_s-4)/3\approx -1$ (equilateral shape). Again, we recover the 
high-peak expression \citep{2010arXiv1006.4763D} if we take the limit 
$\nu\to\infty$. 

\subsubsection{Beyond the high-peak limit}

\reffig{emlb} shows the ratio of the non-Gaussian linear bias correction
arising from the statistics of thresholded regions to that obtained in the
high-peak approximation.  In the literature, $c_N$ is conventionally
replaced with $b_N$, so that this ratio becomes $b_2\s_{0s}^2/(b_1\d_c)$ 
for a primordial bispectrum, and $b_3\s_{0s}^4/(b_1\d_c^2)$ for a primordial 
trispectrum.  Note that these quantities do not depend on the shape of the 
polyspectrum considered. The results are shown at $z=0$ using the 
Gaussian bias factors $b_N$ derived from the Sheth-Tormen mass function 
\cite{1999MNRAS.308..119S,2002MNRAS.329...61S} with $p=0.3$ and $q=0.75$, 
via the PBS approach.  These predictions are clearly at odds with the
simulation results: firstly, for the local bispectrum shape with constant
$\fnl$, there is no evidence of a large suppression relative to the 
high-peak expression \cite{2009MNRAS.396...85D,2009MNRAS.398..321G,
2010MNRAS.402..191P,2010PhRvD..81f3530G} (the correction factor of 
$\sim 0.8$ advocated by \cite{2009MNRAS.398..321G,2010MNRAS.402..191P} 
likely applies for friends-of-friends halo finders solely; see 
\cite{2010arXiv1006.4763D} for a discussion). Secondly, the simulation 
studies of \cite{2010arXiv1010.3722S,2011arXiv1102.3229W} 
unambiguously show that the correction to the high-peak expression depends 
on the shape of the bispectrum. Thirdly, while the suppression seen in 
\reffig{emlb} for $M\gtrsim 10^{14}\hmsun$ is qualitatively consistent with 
that measured for the $\gnl\phi^3$ model for highly biased halos 
\cite{2010PhRvD..81b3006D}, 
the sharp upturn below $\sim 10^{13}\hmsun$ is inconsistent with the findings 
of \cite{2010PhRvD..81b3006D} at high significance.  
This appears to exclude the statistics of thresholded regions and, more 
generally, local bias expansions of the form \refeq{localbias} as a viable 
framework to calculate non-Gaussian bias corrections, at least for realistic 
halo masses ($\lesssim 10^{15} M_{\odot}$).  We return to these issues
in \refsec{PBS-thr}.


\section{Peak-background split: separation of scales}
\label{sec:GPBS}

In this Section, we present our second derivation of the non-Gaussian, 
scale-dependent halo bias, based on the peak-background split argument.  
We show that the fact that the cumulants of the density field depend on the 
smoothing scale $R_s$ induces an important and previously overlooked 
contribution to the non-Gaussian bias correction.

In this approach, we make a separation of scales and split all 
perturbations $\d$, $\phi$, etc. into their long-wavelength (subscript 
``$L$'') and short-wavelength (subscript ``$S$'') pieces, e.g.
\be
\d = \d_L + \d_S,\quad \phi = \phi_L + \phi_S, \quad\dots
\label{eq:pbs}
\ee
Here, short wavelengths signify the scales which impact halo formation 
($\lesssim 10-100\Mpch$), while long wavelengths correspond to the 
scales on which we would like to measure the clustering of halos 
($\gtrsim 100\Mpch$).  
For a Gaussian density field with independent Fourier modes, 
the $L$ and $S$ pieces are statistically independent. In the presence of 
non-Gaussianity, this is no longer the case.  As we will see
shortly, it will be convenient to apply the peak-background split to the
Gaussian primordial curvature perturbation $\phi$.  This approach isolates
the effect of mode-coupling introduced by primordial non-Gaussianity,
allowing for direct physical insights.  
To avoid confusion, we shall denote the physical, non-Gaussian density 
field by $\hat\d$, to distinguish it from the Gaussian density field $\d$ 
related to the Gaussian potential $\phi$.  

\subsection{General cubic non-Gaussianity}
\label{sec:gnlGPBS}

Consider the case of weakly non-Gaussian potential perturbations described 
via non-zero three- and four-point functions.  We can capture the 
non-Gaussian corrections by generalizing the cubic local ansatz 
\refeq{localNG} in Fourier space:
\ba
\Phi(\vk) =\:& \phi(\vk) 
+ \fnl \int\!\!\frac{d^3k_1}{(2\pi)^3}\int\!\!\frac{d^3k_2}{(2\pi)^3} 
\,\wt(\vk_1,\vk_2) \phi(\vk_1)\phi(\vk_2) \d_D(\vk-\vk_{12})\vs 
& + \gnl \int\!\!\frac{d^3k_1}{(2\pi)^3}\int\!\!\frac{d^3k_2}{(2\pi)^3}
\int\!\!\frac{d^3k_3}{(2\pi)^3} 
\,\wth(\vk_1,\vk_2,\vk_3) \phi(\vk_1)\phi(\vk_2)\phi(\vk_3) 
\d_D(\vk-\vk_{123})\;,
\label{eq:phiNG}
\ea
where $\vk_{12\dots} = \vk_1+\vk_2+\dots$, and the two kernels $\wt$, $\wth$
are related to the three- and four-point function, respectively \cite{fsmk}.  
The relation is in general ambiguous, i.e. different kernels can yield the 
same three- and four-point functions.  However, the large-scale limit of the
non-Gaussian bias depends on the squeezed limit of the $N$-point functions,
as we have seen in \refsec{MLB}.  In this limit, the kernels $\wt$, $\wth$
are unique \footnote{This is true excluding unviable kernels that yield large
loop corrections to the power spectrum $P_\phi$.}.  

One possible choice of kernels, which has the nice property (for 
analytical calculations) of being fully symmetric, is
\ba
\wt(\vk_1,\vk_2) =\:& \frac{1}{2\fnl}
\frac{\xi_\Phi^{(3)}(\vk_1,\vk_2,\vk_3)}
{P_1 P_2 + 2\:{\rm perm.}}
\label{eq:wt}\\
\wth(\vk_1,\vk_2,\vk_3) =\:& \frac{1}{6\gnl}
\frac{\xi_\Phi^{(4)}(\vk_1,\vk_2,\vk_3,\vk_4)}
{P_1 P_2 P_3 + 3\:{\rm perm.}}\;,
\label{eq:wth}
\ea
where in the first line, $k_3=|\vk_{12}|$, while in the second line,
$k_4=|\vk_{123}|$, and $P_i \equiv P_\phi(k_i)$.  
We have pulled out the coefficients $\fnl$ and $\gnl$ for convenience.  
Note that, in general, the four-point function also contains terms of 
order $\fnl^2$, which we assume to be included in $\xi^{(4)}_\Phi$ even 
though we parameterize the amplitude by a single coefficient $\gnl$.  

\refeq{wt} is analogous to the kernel $\widetilde{W}(\vk_1,\vk_2)$ defined 
in \cite{fsmk}, and \refeq{wth} is the straightforward generalization to 
the cubic case.  Note that we define the kernels in terms of $\phi$ here, 
while they are defined in terms of $\phi_0(k)\equiv T(k)\phi(k)$ in 
\cite{fsmk}.  The final result (in the large-scale limit) is independent 
of this choice of kernel, which yields $\wt = \wth = 1$ for the local
model.

In the next subsection, we first calculate the effect of long-wavelength
perturbations $\phi_L,\;\d_L$ on the statistics of the small-scale
density field $\d_S$.  We then derive expressions for the
non-Gaussian halo bias for general cubic non-Gaussianity.

\subsubsection{Effect of long-wavelength perturbations on the density field}

We begin by applying the separation of scales, \refeq{pbs}, to \refeq{phiNG}.  
Clearly, for the quadratic part we will obtain the combinations $(SS)$,
$(SL)$, and $(LL)$, while the cubic part yields $(SSS)$, $(SSL)$, $(SLL)$,
$(LLL)$.  The terms involving $L$ solely do not influence halo abundance
(since they do not contribute significantly to the moments of the 
small-scale density 
field).  The terms involving $S$-perturbations only increase the variance, 
skewness, and kurtosis of the small-scale density field. They may thus affect 
the abundance of halos. However, they do so in a scale-independent way and, 
thereby, induce at most a scale-independent bias correction. Hence, in order 
to derive the (scale-dependent) effect of non-Gaussianity on halo clustering, 
we only need to retain the mixed terms.  

We now want to derive an expression for the non-Gaussian small-scale
density field $\hat\d_S(k) = \Mm(k) \Phi_S$.  We obtain it by 
multiplying the short-wavelength part of \refeq{phiNG} by $\Mm(k)$. Next, 
we apply a trick, noting that $\Mm(k)\propto k^2$, and 
$k^2 = [\vk_1 + (\vk-\vk_1)]^2$.  Thus, 
\be
\Mm(k) = \Mm(k_1) + \Mm(|\vk-\vk_1|) 
+ \mathcal{O}(\vk_1\cdot[\vk-\vk_1])\;.  
\label{eq:Mm2}
\ee
When inserting \refeq{Mm2} into the first line of \refeq{phiNG}, we see 
that for the local model, where $\o^{(2)}=1$, the last term in \refeq{Mm2} 
corresponds to the Fourier transform of $\nabla\phi_L\cdot\nabla\phi_S$ 
(recall that we are only dealing with mixed terms).  
When averaging over a region where $\nabla\phi_L$ is approximately a
constant gradient, this term vanishes since $\phi_S$ is uncorrelated with 
$\phi_L$ (see also \cite{2008PhRvD..78l3507A} for a different procedure
in the local case).  Below we will perform precisely such an averaging 
procedure. A similar reasoning can be applied to the non-local case. Hence 
we will drop this term and its analogs in the cubic part of \refeq{phiNG}.  
Note that we have neglected the $k$-dependence of the transfer function
here.  One can circumvent this by defining the kernel in terms of $\phi_0$,
as done in \cite{fsmk}.  \refeq{Mm2} and its generalization to several
$k_i$ then lead to
\ba
\hat\d_S(\vk) =\:& \d_S(\vk) + 2\fnl 
\int\!\!\frac{d^3k_1}{(2\pi)^3} \wt(\vk_1,\vk-\vk_1)
\left[\d_L(\vk_1)\phi_S(\vk-\vk_1) + \d_S(\vk_1)\phi_L(\vk-\vk_1)\right]\vs
& + 3\gnl \int\!\!\frac{d^3k_1}{(2\pi)^3}\int\!\!\frac{d^3k_2}{(2\pi^3)}
\wth(\vk_1,\vk_2,\vk-\vk_{12}) \left [2 \phi_L \delta_L \phi_S
+ \phi_L\phi_L \d_S + 2 \phi_L \phi_S \d_S 
+ \d_L \phi_S\phi_S\right]\;.
\label{eq:dsng}
\ea
In the second line, we have omitted the arguments of $\phi$, $\d$ for brevity 
(the factors in each product are evaluated at $\vk_1$, $\vk_2$, and 
$\vk-\vk_{12}$, respectively).  

In the presence of non-Gaussianity, the statistical properties of $\hat\d_S$ 
can be derived straightforwardly from \refeq{dsng} by taking advantage of the 
fact that $\phi_S,\;\d_S$ are \emph{Gaussian} fields.  We will consider
a region of ``intermediate'' size $R \gg R_s$ over which the long-wavelength
perturbations can be considered constant.  This approximation will break down
when predicting the clustering on scales which contribute significantly
to $\s_{0s}$ (see the discussion below and in \cite{fsmk}).  
We then calculate the variance and skewness of $\hat\d_S$ in the presence of 
``external'' perturbations $\phi_L,\;\d_L$.  To compute the variance 
for instance, we calculate $\<\hat\d_S(\vk)\hat\d_S(\vk')\>$ and integrate 
over $\vk$.  It is sufficient to consider a single (for the quadratic 
terms) or two independent (for the cubic terms) long-wavelength Fourier modes 
and, hence, omit the integrals over $k_1,\;k_2$.  This is because we 
will eventually take derivatives with respect to single long-wavelength 
Fourier modes in order to derive the non-Gaussian scale-dependent bias.  
The variance on scale $R_s$ reads
\ba
\label{eq:sRng}
\hat\s_{0s}^2 \equiv \<\hat\d_{S,R_s}\hat\d_{S,R_s}\>_R 
=\:& \s_{0s}^2 + 4\fnl\left[\phi_L(\vk) \s_{\o s}^2(k)
+ \d_L(\vk) \s_{\o\phi s}^2(k)\right]\vs 
& + 6\gnl \phi_L(\vk_1)\phi_L(\vk_2) \s_{\o s}^2(\vk_1,\vk_2) 
+ 6\gnl \left[\phi_L(\vk_1)\d_L(\vk_2)+\phi_L(\vk_2)\d_L(\vk_1)\right] 
\s_{\o\phi s}^2(\vk_1,\vk_2)
\;,
\ea
where  $\<\cdot\>_R$ indicates an average over a given intermediate 
region of size $R$.  Note
that the terms from quadratic non-Gaussianity are linear in $\phi_L,\;\d_L$,
while those from cubic non-Gaussianity are quadratic in $\phi_L,\;\d_L$.  
For \refeq{sRng}, we have defined the following $k$-dependent spectral 
moments (not to be confounded with \refeq{sigmaAR}, which does not depend
on $k$):
\ba
\s_{\o s}^2(k) \equiv\;& \int\!\!\frac{d^3k_s}{(2\pi)^3}\,\wt(\vk,\vk_s)
\Mm_s^2(k_s)\, P_\phi(k_s)\;\label{eq:sigmao}\\
\s_{\o\phi s}^2(k) \equiv\;& \int\!\!\frac{d^3k_s}{(2\pi)^3}\,\wt(\vk,\vk_s)
\Mm_s(k_s)\, P_\phi(k_s)\;\label{eq:sigmapo}\\
\s_{\o s}^2(\vk_1,\vk_2) \equiv\;& \int\!\!\frac{d^3k_s}{(2\pi)^3}\,\wth(\vk_1,\vk_2,\vk_s)
\Mm_s^2(k_s)\, P_\phi(k_s)\;\label{eq:sigma3o}\\
\s_{\o\phi s}^2(\vk_1,\vk_2) \equiv\;& \int\!\!\frac{d^3k_s}{(2\pi)^3}\,\wth(\vk_1,\vk_2,\vk_s)
\Mm_s(k_s)\, P_\phi(k_s)\;\label{eq:sigma3po}
\ea
In the following, we will ignore the term
$4\fnl\delta_L\s_{\o \phi s}^2$ since it only generates a very small
($\lesssim 10^{-4}\fnl$) scale-independent correction to the halo bias.  

At cubic order in \refeq{dsng}, there are two terms of the type $(LSS)$. 
These terms indicate that the small-scale density acquires a skewness 
(third moment) which is modulated by long-wavelength perturbations. We will 
only include the effect of the first term, $6\gnl \phi_L\phi_S\d_S$, as the 
second term proportional to $\d_L\phi_S^2$ only produces a scale-independent 
correction to the halo bias.  
The three-point function of the small-scale density field induced by
a single long-wavelength perturbation $\phi_L(\vk_l)$ is given by 
\ba
\<\d(\vk)\d(\vk')\d(\vk'')\>_R =\:& 
3\gnl \phi_L(\vk_l)\: (2\pi)^3 \d_D(\vk+\vk'+\vk'')\: \Mm(k)\Mm(k')\Mm(k'') \vs
&\times\left \{ \left [ \wth(\vk_l,\vk',\vk'') + \wth(\vk_l,\vk'',\vk')\right ]
P_\phi(k')P_\phi(k'') + 2\;{\rm perm.}\right\}\;,\label{eq:d3ng}
\ea
where ``2~perm'' indicates the two cyclic permutations of $(k,\;k,'\;k'')$.  
Recall that the subscript $R$ on the expectation value indicates averaging 
over a region where $\phi_L$ is approximately constant.  
In deriving \refeq{d3ng}, we have used that $k = |\vk'+\vk''|$, so that
$\Mm(k')+\Mm(k'')\approx \Mm(k)$ on large scales.  Thus, the three-point
function of the small-scale density field induced by a long-wavelength 
perturbation in cubic non-Gaussianity is equivalent to that arising in a
\emph{quadratic} model of non-Gaussianity described by the effective
three-point function
\be
\xi^{(3)}_{\Phi,\rm eff}(\vk,\vk',\vk'') = f_{\rm NL,eff}(\vk_l)
\left \{\left [ \wth(\vk_l,\vk',\vk'') + \wth(\vk_l,\vk'',\vk')\right ]
P_\phi(k')P_\phi(k'') + 2\;{\rm perm}\right\},
\label{eq:xi3eff}
\ee
where $f_{\rm NL,eff} = 3\gnl \phi_L(\vk_l)$.  Note that 
$\xi^{(3)}_{\Phi,\rm eff}$
generally depends on the scale $k_l$ of the long-wavelength perturbation.  
We can now calculate the skewness parameter of the small-scale non-Gaussian 
density field, taking out the scaling with the long-wavelength mode $\phi_L$: 
\ba
{\hat S}^{(3)}_s \equiv\:& 
\frac{\<\hat\d_{s,R_s}^3\>_R}{\<\hat\d_{s,R_s}^2\>_R^2}
= 3\gnl \phi_L(\vk_l) S^{(3)}_{\o s}(k_l),\label{eq:S3ng}\\
S^{(3)}_{\o s}(k_l) \equiv\:&
\frac{6}{\s_{0s}^4}\int\!\!\frac{d^3k_1}{(2\pi^3)} \Mm_s(k_1) P_\phi(k_1)
\int\!\!\frac{d^3k_2}{(2\pi^3} \Mm_s(k_2) P_\phi(k_2) \vs
&\quad\quad\times \wth(\vk_l,\vk_1,\vk_2) \Mm_s(|\vk_1+\vk_2|)\;.
\label{eq:So}
\ea
Here, we have noted that $\<\hat\d_s^3\>$ is already linear in $\gnl$,
so that we can set $\<\hat\d_s^2\>=\s_{0s}^2$.  

Summarizing, the effect of long-wavelength modes in general cubic
non-Gaussianity is to rescale the local small-scale variance of 
the density field [\refeq{dsng}], as was discussed for the quadratic case in 
\cite{2008PhRvD..77l3514D,2008JCAP...08..031S,fsmk}.  This rescaling
is linear in the long-wavelength modes for the quadratic ($\fnl$) term, and
quadratic in $\d_L,\;\phi_L$ for the cubic ($\gnl$) term.  
The terms quadratic in the $L$-modes induce a non-Gaussian correction 
to the second order bias $\bt$.  We will not consider this correction
here as it does not significantly impact the halo power spectrum.  
Furthermore, a long wavelength mode in a cubic model also induces a local 
three-point function (skewness) in the density field [\refeq{d3ng}]: observers 
in a region with $\phi_L \neq 0$ see a local Universe with an effective 
quadratic non-Gaussianity described by the ``primordial'' three-point function
$\xi^{(3)}_{\Phi,\rm eff}$ [\refeq{xi3eff}].  

\subsubsection{Non-Gaussian corrections to the linear bias}
\label{sec:Dbgnl}

Let us now consider the halo abundance $\hat{n}_h(\vx)$ in some region of size 
$R$, with $R_s\ll R\ll R_l$, and $R_s$ being the Lagrangian scale associated
with a halo mass $M$.  Throughout, we will assume that $\hat{n}_h$
depends only on the matter density $\rho_R$ averaged over $R$, and the 
moments of the small-scale fluctuations: $\hat{\sigma}_{0s}^2$, 
$\hat{S}_s^{(3)}$, $\cdots$.  
While, in the Gaussian case, a perturbation $\d_L$ only changes the average 
density [$\rho_R\to \rho_R(1+\d_L)$], it also affects all the cumulants of
the density field when the initial conditions are non-Gaussian.
Applying the chain rule, we find
\begin{align}
\label{eq:bI_pbs_first}
\bo(k)
\equiv
\frac{1}{\hat{\bar{n}}_h} \frac{d \nhh}{d\d_L(k)}
\biggr\lvert_{\d_L=0}
=
\frac{\partial \ln \nhh}{\partial \ln\rho_R}
+
\frac{\partial \ln \nhh}{\partial \ln\hat{\sigma}_{0s}}
\frac{\partial\ln\hat{\sigma}_{0s}}{\partial \delta_L(k)}
+
\frac{\partial \ln \nhh}{\partial \hat{S}_{s}^{(3)}}
\frac{\partial\hat{S}_s^{(3)}}{\partial\delta_L(k)}
+\cdots.
\end{align}
Here, $\nhh$ is the average number density of halos of mass $M$ with 
non-Gaussian initial conditions, and all derivatives are evaluated at 
$\d_L=0$.  Owing to isotropy, $\bo(k)$ only depends on the magnitude of 
the $k$-vector.  
The first term in \refeq{bI_pbs_first} is the usual Gaussian bias $b_1$, 
while the second and third terms yield the non-Gaussian corrections.  
Thus, the non-Gaussian contribution $\D\bo(k)$ to the linear bias $\bo(k)$ 
arises from the dependence of the halo abundance on the variance 
and skewness of the density field.

Let us deal with the variance first.  As we have seen in the last 
Section, the change in the variance from cubic non-Gaussianity is 
$\mathcal{O}(\d_L^2)$.
Hence, these terms do not contribute to the \emph{linear} bias and 
\refeq{dsng} gives
\be
\frac{\partial\ln\hat\s_{0s}}{\partial\d_L(k)}\biggr\lvert_{\d_L=0} 
= 2\fnl \frac{\s_{\o s}^2(k)}{\s_{0s}^2} \Mm^{-1}(k).
\label{eq:dsddL}
\ee
Note that this expression in general depends on the smoothing scale $R_s$
or, equivalently, the halo mass $M$.  

To proceed further, we will restrict ourselves to the case of a 
universal mass function  for Gaussian initial conditions. 
Therefore, the Gaussian halo number density is given by \refeq{umassfn}.  
Throughout this Section, we will not need to specify $f(\nu)$ explicitly.  
The non-Gaussian halo abundance $\nhh$ will thus depend on $\hat\s_{0s}$ 
through the significance $\nu=\d_c/\hat\s_{0s}$ and the Jacobian 
$\partial\hat\s_{0s}/\partial\ln M$. Noting that the Gaussian bias is 
$b_1 = -\d_c^{-1}d\ln f(\nu)/d\ln\nu$, and taking the derivative of 
\refeq{dsddL} with respect to $\ln M$, we obtain
\ba
\frac{\partial\ln\nhh}{\partial\ln\s_{0s}}
\frac{\partial\ln\hat\s_{0s}}{\partial\d_L(k)} \biggr\lvert_{\d_L=0}
=\:& 2\fnl\Mm^{-1}(k) \frac{\s_{\o s}^2(k)}{\s_{0s}^2}
\bigl[ b_1 \d_c + 2\eps_{\o s}(k) \bigr]\;,
\label{eq:boterm}\\
\eps_{\o s}(k) \equiv\:& 
\frac{\partial\ln\s_{\o s}^2(k)}{\partial\ln\s_{0s}^2} - 1\;.
\label{eq:eos}
\ea
The second term in the square brackets, $2\eps_{\o s}(k)$, has previously 
been neglected \cite{fsmk,2010arXiv1010.3722S}.  
It vanishes in the scale-independent local model, for which $\wt=1$ and
$\s_{\o s}=\s_{0s}$, but is non-zero and generally significant for other 
bispectrum shapes.  Physically, this term comes about because a 
scale-dependent rescaling of the variance [\refeq{dsddL}] also changes the 
significance interval $d\nu$ that corresponds to a fixed mass interval $dM$.  
This in turn affects the abundance of halos at a fixed mass and 
thus contributes to the non-Gaussian bias.  
The term is absent in the results of the thresholding approach 
(\refsec{MLB}), since the cumulative two-point correlation 
$\xnu(r)$ is computed at a fixed smoothing scale $R_s$.  We return to this
point in \refsec{PBS-thr}.  

In order to derive the effect of cubic non-Gaussianity, we need to determine 
the dependence of $\nhh$ on $\hat{S}^{(3)}_s$, i.e. the effect of a
primordial three-point function on the average abundance of halos.  
Different (albeit related) expressions have been proposed
for the change in the halo abundance induced by primordial non-Gaussianity
\citep{2000ApJ...541...10M,2008JCAP...04..014L,2009MNRAS.398.2143L,
2010MNRAS.405.1244M,2010A&A...514A..46V,2010arXiv1005.1203D,
2011arXiv1102.1439L,2011arXiv1103.2586Y}. For definiteness, we will adopt 
the prescription of \cite{2008JCAP...04..014L} derived from an Edgeworth 
expansion of $P_1(>\nu)$ (see also \refsec{CPBS}),
\begin{align}
\nhh(\rho,\hat\s_{0s},\hat{S}^{(3)}_s) &= \nh(\rho,\hat\s_{0s},0)
\Biggl[ 1 + \frac{1}{6}\hat\s_{0s}\hat{S}^{(3)}_s(\nu^3-3\nu) 
+ \frac{1}{6}\frac{\partial(\hat\s_{0s}\hat{S}^{(3)}_s)}
{\partial\ln\hat\s_{0s}}\left(\nu-\frac{1}{\nu}\right)\Biggr] \nonumber \\
&= \nh(\rho,\hat\s_{0s},0)
\Biggl\{ 1 + \frac{1}{6}\hat\s_{0s}\hat{S}^{(3)}_s
\Biggl[\left(\nu^3-3\nu\right)+\left(1+\frac{\partial\ln\hat{S}^{(3)}_s}
{\partial\ln\hat\s_{0s}}\right)\left(\nu-\frac{1}{\nu}\right)\Biggr]\Biggr\}\;.
\label{eq:nhhS3}
\end{align}
In principle however, any other prescription for the response of halo number 
counts to a small-scale skewness of the density field could be employed here. 
From \refeq{nhhS3}, we derive
\begin{align}
\nonumber
\frac{6}{\s_{0s}^2}\frac{\partial\ln\nhh}{\partial\hat{S}^{(3)}_s}
\biggr\lvert_{\d_L=0} &= \frac{1}{\s_{0s}} 
\Biggl[ \left(\nu^3-3\nu\right) + 
\left(1 + \frac{\partial\ln\hat{S}^{(3)}_s}{\partial\ln\hat\s_{0s}}\right)
\Biggr\lvert_{\d_L=0} \left(\nu-\frac{1}{\nu}\right)\Biggr] \\
&= b_2 \d_c +
\left(1 + \frac{\partial\ln\hat{S}^{(3)}_s}{\partial\ln\s_{0s}}\right)
b_1 \;.
\label{eq:dnhhdS3}
\end{align}
In the last equality, we have identified the $\nu$-polynomials with the 
Gaussian peak-background split biases derived from the multiplicity function 
$f(\nu)=\sqrt{2/\pi}\,\nu\exp(-\nu^2/2)$, since our parameterization of 
$\nhh$ in terms of $\hat{S}^{(3)}$ was derived within the  Press-Schechter 
formalism \cite{2008JCAP...04..014L}.  
While for high peaks $\nu \gg 1$, the first term in the last equality will 
dominate, for more abundant halos the second term can contribute 
significantly.  The latter again arises because of the dependence of 
$S_{\o s}^{(3)}$ on the smoothing scale $R_s$.  

Finally, using \refeq{S3ng} we find
\be
\frac{\partial\hat S^{(3)}_s}{\partial\d_L(k)} \biggr\lvert_{\d_L=0}
= 3\gnl \Mm^{-1}(k) S^{(3)}_{\o s}(k).
\ee
Then, using \refeq{bI_pbs_first} together with Eqs.~\refe{boterm} and 
\refe{dnhhdS3}, we can assemble the expression for the scale-dependent halo 
bias in a general, cubic order model of non-Gaussianity:
\ba
\D\bo^{(\kappa)}(k) =\:& 2\fnl\Mm^{-1}(k) \frac{\s_{\o s}^2(k)}{\s_{0s}^2}
\bigl[ b_1 \d_c + 2\eps_{\o s}(k) \bigr]\vs
& + \frac{1}{2}\gnl \Mm^{-1}(k) \s_{0s}^2 S^{(3)}_{\o s}(k) 
\left [b_2 \d_c +
\left(1 + \frac{\partial\ln S^{(3)}_{\o s}(k)}{\partial\ln\s_{0s}}\right)b_1
\right]
\label{eq:Dbo-general}
\ea
The superscript $(\kappa)$ emphasizes that this correction is $k$-dependent,
and distinguishes it from a $k$-independent non-Gaussian bias which we shall 
denote with a superscript $(\iota)$.  Note that the terms in the first line of
\refeq{Dbo-general} apply for any universal mass function prescription.  
On the other hand, the coefficients in the square brackets of the
second line will change if a different prescription for 
$\partial\ln\nhh/\partial \hat S^{(3)}_s$ is adopted.

\subsection{Application to models of non-Gaussianity}

\subsubsection{Local non-Gaussianity}
\label{sec:pbslocal}

In the local model [\refeq{localNG}], the kernels \refeqs{wt}{wth} are 
simply $\wt=\wth=1$. Thus, $\s_{\o s}\to \s_{0s}$, and the skewness 
$S^{(3)}_{\o s}$ induced by a long-wavelength perturbation becomes 
$\Sk_{s,\rm loc}$, i.e. the skewness in a local quadratic model with 
$\fnl=1$ [\refeq{S3loc}].  A more direct way to derive this result is to note
that the $(LSS)$ terms in the second line of \refeq{dsng} 
are obtained by applying the Poisson equation to an effective 
non-Gaussian potential
\be
\hat\phi_S = \phi_S + (3\gnl \phi_L) \phi_S^2\;.
\ee
This relation tells us that, in the presence of cubic local non-Gaussianity, 
a region with a long-wavelength perturbation $\phi_L$ looks like a Universe 
with a local quadratic $\fnl=3\gnl\phi_L$.  

Since $\eps_{\o s}=0$, the correction to the first order bias 
\refeq{Dbo-general} then simplifies to 
\be
\D\bo^{(\kappa)}(k) = \left [ 2\fnl b_1 \d_c
+ \frac{1}{2}\gnl\s_{0s}^2 \Sk_{s,\rm loc}\: \epsilon_S \right]
\Mm^{-1}(k)\;,
\label{eq:Db1}
\ee
where we have defined
\begin{align}
\epsilon_S \equiv\:& b_2 \d_c +
\left(1 + \frac{\partial\ln\Sk_{s,\rm loc}}{\partial\ln\s_{0s}}\right)b_1\;.
\label{eq:eS}
\end{align}
The term linear in $\fnl$ recovers the well-known result for the local 
quadratic model (this is due to the fact that 
$\hat\s_{0s}/\s_{0s} = 1+2\fnl\phi_L$ is scale-independent).  
However, the term linear in $\gnl$ departs from the high-peak expression 
derived in \cite{2010PhRvD..81b3006D} as it includes a correction involving 
the logarithmic slope of $\Sk_{s,\rm loc}$ on $\s_{0s}$.  We will
return to this point in \refsec{PBS-thr}.  
In the range $R_s\sim 1-10\hmpc$, the scale dependence of $\Sk_{s,\rm loc}$ 
is accurately reproduced by an empirical power-law relation,
$\Sk_{s,\rm loc}\approx 3.08\times 10^{-4} \hat\s_{0s}^{-0.855}$ for our 
fiducial cosmology (this agrees with the findings of 
\cite{2010ApJ...724..285C,2010arXiv1012.2732E}). Hence, the second term 
in \refeq{eS} is approximately $0.145\,b_1$ and, therefore, not negligible. 

\begin{figure*}
\center
\resizebox{0.49\textwidth}{!}{\includegraphics{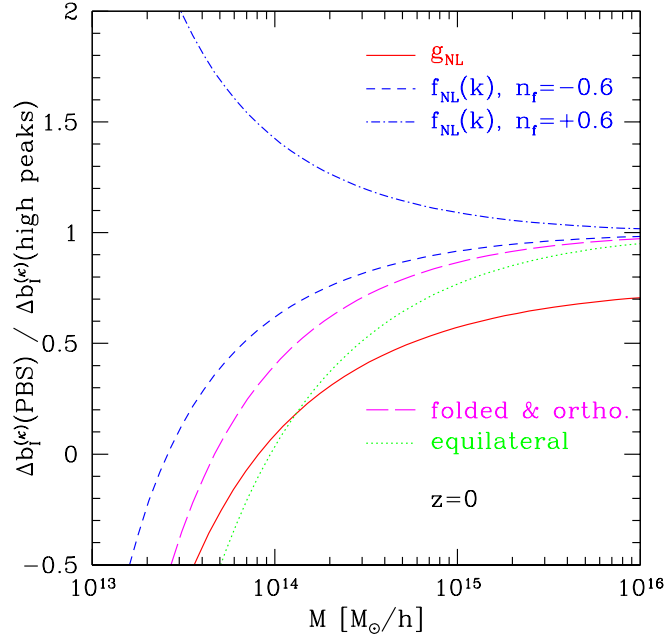}}
\caption{Ratio of the non-Gaussian correction to the linear bias predicted
by the peak-background split approach to that obtained in the high-peak
limit. Results are shown at $z=0$ as a function of the halo mass $M$ for a 
local trispectrum with cubic parameter $\gnl$ (solid curve), a 
local bispectrum with $k$-dependent quadratic parameter $\fnl$ and index
$n_f=\pm 0.6$ (dashed and dot-dashed curve), the folded and orthogonal 
template (long-dashed curve) and the equilateral bispectrum shape
(dotted curve). In contrast to \reffig{emlb}, the ratio sensitively depends 
on the shape of the primordial $N$-point function.} 
\label{fig:epbs}
\end{figure*}

The ratio of the peak-background split prediction to the high-peak result 
is given 
by $\hat\s_{0s}^2\epsilon_S/(b_1 \d_c^2)$. In \reffig{epbs}, the value of 
this ratio in the limit $k\to 0$ is shown as the solid curve. We assume a 
critical collapse density $\d_c=1.69$ and, in the calculation of the 
Gaussian biases $b_N$, we employ again a Sheth-Tormen 
multiplicity function with $p=0.3$ and $q=0.75$. As can be seen, the ratio 
depends strongly on the halo mass $M$. At the redshift assumed here ($z=0$), 
it reverses sign around $M\simeq 7\times 10^{13}\hmsun$. 

\subsubsection{Scale-dependent and non-local non-Gaussianity}
\label{sec:pbsfnl}

We now turn to the other models of primordial non-Gaussianity introduced 
in \refsec{prelim}.  Since these are all quadratic models, we can ignore
the term linear in $\gnl$ in \refeq{Dbo-general}.  The dependence of 
$\D\bo(k)$ on the shape of non-Gaussianity enters through the moment 
$\s_{\o s}(k)$ [\refeq{sigmao}] and through the new correction 
proportional to $\partial\ln\s_{\o s}(k)/\partial\ln\s_{0s}$. Since we 
are interested in the large-scale limit, we can use the scaling of the 
kernel $\wt$ in the squeezed limit in order to simplify the analytical
expressions.  
For the local quadratic model with $k$-dependent $\fnl$ (see \refeq{Block}), 
the kernel in this limit reduces to
\be
\o(\vk,\vk_s-\vk) \stackrel{k\rightarrow 0}{=} 
\left(\frac{k_s}{k_p}\right)^{n_f}.
\ee
As a result,
\be
\s_{\o s}^2(k) = \frac{\s_{\a s}^2}{k_p^{n_f}} ,
\ee
where $\a \equiv n_f/2$ and the spectral moment $\s_{\a s}$ is defined in
\refeq{sigmaAR} with $R=R_s$.  Using the first line of \refeq{Dbo-general}, the
non-Gaussian, scale-dependent bias correction is then given by
\be
\label{eq:DbIkfnl}
\D b_{\rm I,sc.loc.}^{(\kappa)}(k) = 2\fnl(k_p) k_p^{-n_f}
\left(\frac{\s_{\a s}}{\s_{0s}}\right)^2
\left[
b_1 \d_c + 2\left(\frac{\partial\ln \s_{\a s}}{\partial\ln\s_{0s}}-1\right)
\right]
\Mm^{-1}(k)\;, \qquad \a = n_f/2\;.
\ee
The logarithmic derivative of $\s_{\a s}$ w.r.t. $\s_{0s}$ is always larger 
(smaller) than unity for $\a>0$ ($\a<0$), and reaches unity for $\a\neq 0$ 
only in the limit $\s_{0s}\to 0$.  For the folded and orthogonal bispectrum 
shapes [\refeq{Bfol} and \refeq{Bort}], the kernel asymptotes to \cite{fsmk}
\be
\o(\vk,\vk_s-\vk) \stackrel{k\rightarrow 0}{=} \frac{3}{2}A
\left(\frac{k_s}{k}\right)^{2\a},
\ee
with $2\a = (n_s-1)/3 - 1 = (n_s-4)/3 \approx -1$, and $A=1$ (folded) or 
$A=-2$ (orthogonal). Note that we have neglected corrections of order 
$(k/k_s)^2$ and higher here (although they are easy to include in a numerical
calculation).  Inserting this result into \refeq{Dbo-general}, 
we arrive at
\be
\label{eq:DbIfolded/ort}
\D b_{\rm I,fol/ort}^{(\kappa)}(k) = 3A\fnl 
\left(\frac{\s_{\a s}}{\s_{0s}}\right)^2
\left [
b_1 \d_c + 2\left(\frac{\partial\ln \s_{\a s}}{\partial\ln \s_{0s}}-1\right)
\right] 
k^{-2\a} \Mm_s^{-1}(k)\;,\qquad \a = (n_s-4)/6\;.
\ee  
Finally, for the equilateral bispectrum, we have
\be
\o(\vk,\vk_s-\vk) \stackrel{k\rightarrow 0}{=} 3
\left(\frac{k_s}{k}\right)^{2\a},
\ee
with $2\a = 2(n_s-4)/3 \approx -2$, which in close analogy with the 
folded case leads to
\be
\D b_{\rm I,eql}^{(\kappa)}(k) = 6\fnl \left(\frac{\s_{\a s}}{\s_{0s}}\right)^2
\left [
b_1 \d_c + 2\left(\frac{\partial\ln \s_{\a s}}{\partial\ln \s_{0s}}-1\right)
\right] 
k^{-2\a} \Mm_s^{-1}(k)\;,\qquad \a = (n_s-4)/3\;.
\label{eq:DbIeql}
\ee
Our results agree with those of \cite{2010arXiv1010.3722S} (for the 
$k$-dependent $\fnl$) and \cite{fsmk,2010arXiv1006.4763D} (for the folded 
and equilateral shapes) apart from a factor $\epsilon_\a/(b_1\d_c)$, where
\be
\epsilon_\a\equiv
b_1\d_c + 2\left(\frac{\partial\ln \s_{\a s}}{\partial\ln \s_{0s}}-1\right)
\;.
\label{eq:eA}
\ee
This quantity clearly depends on the shape of primordial non-Gaussianity
through the slope $\a$.  
The ratio $\epsilon_\a/(b_1\d_c)$, which quantifies the deviation from 
the high-peak approximation, is shown in \reffig{epbs} for two $k$-dependent 
$\fnl$ models with $n_f=\pm 0.6$ (dashed and dot-dashed curve), for the
folded and orthogonal templates (long-dashed curve), and for the equilateral 
(dotted curve) bispectrum shapes. As can be seen, the ratio of the PBS to 
the high-peak prediction depends strongly on $b_1^{\rm E}$ and the value of 
$\alpha$. It is larger (smaller) than unity when $\a>0$ ($\a<0$). The 
suppression relative to the high-peak prediction is strongest for the 
equilateral bispectrum shape, for which $\a\approx -1$, but significant for 
all bispectrum shapes we considered. Clearly, this strong mass dependence 
could be exploited to help constrain the shape of the primordial 
non-Gaussianity.  As shown in \cite{2011PhRvD..84f1301D}, the results of 
N-body simulations match the predictions derived in this section well.


\section{Peak-background split: conditional mass function}
\label{sec:CPBS}

In this Section, we consider the third derivation of the non-Gaussian 
bias based on the conditional halo mass function.  This is essentially a 
peak-background split approach since we again consider the effect of 
adding a background perturbation $\d_l$ of characteristic wavelength 
$R_l\gg R_s$ on the number density $\nh$ of biased tracers.  However, in 
contrast to the previous formulation, we consider a background density 
perturbation $\d_l$ which is statistically correlated with small-scale 
density fluctuations. 
As shown in \cite{2010PhRvD..82j3529D}, such a peak-background split 
approach can be applied to obtain the scale-dependent bias factors of 
(Gaussian) density peaks at all orders. Here, we demonstrate that the 
the implementation of \cite{2010PhRvD..82j3529D} can be generalized to
derive the non-Gaussian bias corrections.

In what follows, $\nhh$ and $\hat P_i$ will denote non-Gaussian number 
densities and probability distributions, whereas $\nh$ and $P_i$ will 
designate the Gaussian quantities. Since we will hereafter deal with the 
non-Gaussian density field, we shall revert to the notation of \refsec{MLB} 
and simply denote the latter as $\d_s,\d_l$ (and $\nu_s,\nu_l$).  
For simplicity, and since we are interested in the behavior on large scales, 
we shall ignore the peak constraint, which leads to corrections scaling as 
$k^2$ and higher powers.  In other words, we will assume that, for Gaussian
initial conditions, the number density of virialized objects $\nh(\nu,R_s)$ 
identified on the scale $R_s$ follows a Press-Schechter mass function.  

\subsection{Cumulants and conditional mass function}

Extending the derivation of the Press-Schechter mass function to the
non-Gaussian case, we start from 
\be
\label{eq:nhng}
\nhh(\nu,R_s)=-2\frac{\rhob}{M}\frac{d\hat P_1(>\nu,M)}{dM}
=-2\frac{\rhob}{M}\frac{d}{dM}\int_{\nu}^\infty\!\!
dx\,\hat P_1(x,R_s)\;,
\ee
where $\hat{P}_1(\nu,R_s)$ is the probability that the linear density contrast
of a Lagrangian region of mass $M\propto R_s^3$ equals $\d_c=\nu\sigma_{0s}$, 
and $\hat P_1(>\nu,M)$ is the probability that the same density contrast
exceeds $\delta_c$.  In this Section, we shall use the parameter
$\nu \equiv \delta_c/\sigma_{0s}$ exclusively for the significance 
corresponding to  the critical density with smoothing $R_s$.  On the other 
hand, $\nu_s \equiv \delta_s/\sigma_{0s}$ and 
$\nu_l \equiv \delta_l/\sigma_{0l}$ stand for $N(0,1)$-distributed 
stochastic variables corresponding to density perturbations on 
small and large scales, respectively.  
As in \refsec{MLB}, we express the non-Gaussian joint probability density 
$P(\vy)$ for the $N$-dimensional vector of variables $\vy$ in terms of
the corresponding Gaussian probability density, 
by using the following general expansion:
\be
\label{eq:png}
\hat P(\vy)=
\exp\left[\sum_{m=3}^\infty\frac{(-1)^m}{m!}\!\sum_{\mu_1\cdots\mu_m}^N
\left\langle y_{\mu_1}\cdots y_{\mu_m}\right\rangle_c \frac{\partial^m}
{\partial y_{\mu_1}\cdots\partial y_{\mu_m}}\right] P(\vy)\;,
\ee
where $\left\langle y_{\mu_1}\cdots y_{\mu_m}\right\rangle_c$ are connected
cumulants and $P(\vy)$ is the multivariate Gaussian distribution 
characterized by the covariances $\left\langle y_{\mu_1}y_{\mu_2}\right\rangle$
\citep[e.g.,][]{2003ApJ...584....1M}. On inserting this expression into
\refeq{nhng}, the non-Gaussian mass function becomes
\begin{align}
\label{eq:P1hat}
\nhh(\nu,R_s) &= -\sqrt{\frac{2}{\pi}}\frac{\rhob}{M}
\frac{d}{dM}\int_{\nu(R_s)}^\infty\!\!dx\,
\exp\Biggl[\sum_{m=3}^\infty\frac{(-1)^m}{m!}\left\langle\nu_s^m\right\rangle_c
\frac{\partial^m}{\partial x^m}\Biggr]e^{-x^2/2} \\
&\approx \sqrt{\frac{2}{\pi}}\,\nu\,e^{-\nu^2/2}
\Biggl(1+\sum_{m=3}^\infty\frac{\left\langle\nu^m_s\right\rangle_c}{m!}
H_m(\nu)\Biggr)\frac{\rhob}{M^2}\frac{d\ln\nu}{d\ln M}
-\sqrt{\frac{2}{\pi}}\frac{\rhob}{M}\int_\nu^\infty\!\!dx\,e^{-x^2/2}
\frac{d}{dM}\Biggl(1+\sum_{m=3}^\infty\frac{\left\langle\nu^m_s\right\rangle_c}
{m!}H_m(x)\Biggr) \nonumber \;.
\end{align}
In the second line we have assumed that all the cumulants are much smaller 
than unity. This formula agrees with that obtained by 
\cite{2008JCAP...04..014L} at first order. Note that the excursion set 
approach yields additional, albeit small corrections to the Press-Schechter 
expressions \cite{2009MNRAS.398.2143L}. However, we will ignore them in 
what follows.

We now calculate the conditional mass function $\nhh(\nu,R_s|\nu_l,R_l)$.  
By definition, the conditional probability for having a small scale 
overdensity $\nu_s$ on scale $R_s$ given a large-scale overdensity $\nu_l$ 
on scale $R_l$ is
\be
\hat P(\nu_s,R_s|\nu_l,R_l)
=
\frac{\hat P_2(\nu_s,R_s,\nu_l,R_l)}{\hat P_1(\nu_l,R_l)}\;.
\ee
The resulting conditional mass function thus is
\be
\nhh(\nu,R_s|\nu_l,R_l)
=
-2\frac{\bar\rho}{M} 
\frac{d}{dM}
\int_{\nu}^\infty
d\nu_s \hat P(\nu_s,R_s|\nu_l,R_l)
=
-2 \frac{\rhob}{M} 
\Bigl[\hat P_1(\nu_l,R_l)\Bigr]^{-1}\frac{d}{dM}
\int_\nu^\infty\!\!dx\,\hat P_2(x,R_s,\nu_l,R_l)\;.
\ee
The joint probability distribution $\hat P_2(\nu,R_s,\nu_l,R_l)$ is readily 
obtained from Eq.(\ref{eq:png}),
\be
\hat P_2(\nu,R_s,\nu_l,R_l)\approx 
\Biggl(1+\sum_{N=3}^\infty\sum_{m=0}^N
\frac{\left\langle\nu_s^m\nu_l^{N-m}\right\rangle_c}{m! (N-m)!}\,
H_{m,N-m}(\nu,\nu_l,\epsilon)\biggr)\frac{f(\nu,\nu_l,\epsilon)}
{2\pi\sqrt{1-\epsilon^2}}\;.
\label{eq:P2hat}
\ee
Here, the correlator stands for
\be
\left\langle\nu_s^m\nu_l^{N-m}\right\rangle_c
=
\s_{0s}^{-m}\s_{0l}^{N-m}
\left\langle\delta_s^m(\bm{x})\delta_l^{N-m}(\bm{x})\right\rangle_c \;,
\ee
where $\bm{x}$ is an arbitrary spatial location. 
The function $f(\nu,\nu_l,\epsilon)$ is the exponential piece of the Gaussian 
bivariate distribution, whereas $H_{mn}(\nu,\nu_l,\epsilon)$ are bivariate 
Hermite polynomials. They can be computed by taking derivatives of 
$f(\nu,\nu_l,\epsilon)$. Namely,
\be
\label{eq:def_Hmn}
(-1)^{m+n}\frac{d^m}{d\nu^m}\frac{d^n}{d\nu_l^n}f(\nu,\nu_l,\epsilon)=
f(\nu,\nu_l,\epsilon)H_{mn}(\nu,\nu_l,\epsilon),\qquad 
f(\nu,\nu_l,\epsilon)\equiv 
\exp\left[-\frac{\nu^2+\nu_l^2-2\epsilon\nu\nu_l}
{2\left(1-\epsilon^2\right)}\right]\;.
\ee
We define mixed spectral moments via 
\be
  \s_{n\times}^2 \equiv \frac{1}{2\pi^2}
  \int_0^\infty\!\!dk\, k^{2(n+1)}\,P_\phi(k)\,{\cal M}_s(k){\cal M}_l(k)\;,
\ee
quantifying the cross-correlation between small and large scales (the 
$\times$ denotes the splitting of smoothing scales: one filter is of size 
$R_s$, the other of size $R_l$).  Further, we define the quantity 
$\Sigma_\times^2$ as
\be
\label{eq:sigmacross}
\Sigma_\times^2\equiv\frac{1}{2\pi^2}\int_0^\infty\!\!dk\, k^2 \Pp(k) 
\Mm_s(k) \Mm_l(k) {\cal S}(k,R_s,R_l)\;,
\ee
where the form factor ${\cal S}$ generally is a function of $k$, $R_s$ and 
$R_l$. This definition is broad enough to describe all the spectral moments 
and the cumulants of the density field.  For instance, setting 
${\cal S}(k,R_s,R_l)=k^2$ yields $\Sigma_\times^2 = \s_{1\times}^2$.  
In the following, we will use the following kernel for $\Sigma_\times^2$:
\be
{\cal S}(k,R_s)\equiv 4\s_{0s}^2 {\cal F}_s^{(N)}(k,X)
{\cal M}_s^{-1}(k)\;.
\label{eq:Skernel}
\ee
Inserting the expression for the form factor \refeq{fshape}, we see
that $\Sigma_\times^2$ becomes
\be
\Sigma_\times^2 = \s_{0s}^{N-1}\s_{0l}\; \<\nu_s^{N-1} \nu_l\>_c\;,
\ee
i.e. a mixed $N$-th order moment of the density field induced by the 
primordial $N$-point function.  

\subsection{Relative overabundance of rare objects}

The non-Gaussian corrections to the $N$-th order Gaussian bias parameters 
$b_N$ can be calculated by expanding the relative overabundance of biased 
tracers $\nhh(\nu,R_s|\nu_l,R_l)/\nhh(\nu,R_s)-1$ at order $\d_l^N$.  
However, throughout the remainder of this Section we will consider only the 
correction to the linear bias. Taking the ratio of the conditional mass 
function to the universal one yields 
\be
\label{eq:dpkng}
\d_h(\d_l)\equiv
\frac{\nhh(\nu,R_s|\nu_l,R_l)}{\nhh(\nu,R_s)}-1 =
\frac{\frac{d}{dM}\int_\nu^\infty\!\!dx\,\hat P_2(x,R_s,\nu_l,R_l)}
{\hat P_1(\nu_l,R_l)\frac{d}{dM}\int_\nu^\infty\!\!dx\,
\hat P_1(x,R_s)}-1\;.
\ee
Now comes a crucial step in the calculation. As $R_l$ increases, the ratio 
$\left\<\nu^m_s\nu_l^{N-m}\right\>/\s_{0l}^2$ (which is the analog of 
$\s_{n\times}^2/\s_{0l}^2$ in the calculation of the peak bias factors) 
remains finite only if the corresponding form factor ${\cal S}(k,R_s,R_l)$ 
does not depend on $R_l$ (again, this applies when expanding to linear order 
in $\nu_l$).  
This implies that, in \refeq{dpkng}, only the terms involving the cumulants 
$\left\<\nu^N_s\right\>_c$ or $\left\<\nu^{N-1}_s\nu_l\right\>_c$ will survive.  
Therefore, upon taking the limit $R_l\to\infty$, we arrive at
\be
\label{eq:dpkng1}
\d_h(\d_l) =
\frac{\frac{d}{dM}\int_\nu^\infty\!\!dx\,
\biggl\{1+\sum_{N=3}^\infty\frac{1}{N!}\Bigl[
\left\<\nu_s^N\right\>_c H_{N,0}(x,\nu_l,\epsilon)
+N\left\<\nu_s^{N-1}\nu_l\right\>_c H_{N-1,1}(x,\nu_l,\epsilon) 
\Bigr]\biggr\}\frac{\exp\left[-\frac{(x-\epsilon\nu_l)^2}{2(1-\epsilon^2)}\right]}
{\sqrt{1-\epsilon^2}}}{\frac{d}{dM}\int_\nu^\infty\!\!dx\,
\Bigl[1+\sum_{N=3}^\infty\frac{\left\<\nu_s^N\right\>_c}{N!} 
H_N(x)\Bigr] e^{-x^2/2}}-1\;.
\ee
In order to calculate the non-Gaussian contribution to $b_1$, 
it is sufficient to expand the right-hand side of \refeq{dpkng1} at 
order $\d_l$. The first term appearing in the square brackets can be
reexpressed as
\begin{align}
\frac{1}{N!}\left\<\nu_s^N\right\>_c H_{N,0}(x,\nu_l,\epsilon)
&= \frac{1}{N!}\left\<\nu_s^N\right\>_c
\left(1-\epsilon^2\right)^{-N/2}H_N\!\left(\frac{x-\epsilon\nu_l}
{\sqrt{1-\epsilon^2}}\right) \\
&\approx \frac{1}{N!}\left\<\nu_s^N\right\>_c\Bigl[H_N(x)
-N\epsilon\nu_l H_{N-1}(x)\Bigr]+{\cal O}(\nu_l^2)\nonumber\;.
\end{align}
In the second line, we successively set $\epsilon\to 0$ (we can ignore 
terms involving $\epsilon^2$) and employed the relation 
$H_N'(x)=N H_{N-1}(x)$ to expand the result at first order in $\nu_l$. 
To simplify the second term in the curly brackets of \refeq{dpkng1}, 
we use the fact that $f(\nu,\nu_l,\epsilon)$ in \refeq{def_Hmn} satisfies 
the following identity
\be
\left[
\frac{\partial }{\partial\nu_l}
+
\epsilon \frac{\partial }{\partial x}
\right]
f(x,\nu_l,\epsilon) = 
-\nu_l 
f(x,\nu_l,\epsilon) \;.
\ee
Therefore, 
\begin{align}
\nonumber
H_{N-1,1}(x,\nu_l,\epsilon)
\equiv&
\frac{(-1)^N}{f(x,\nu_l,\epsilon)}
\frac{\partial^{N-1}}{\partial x^{N-1}}
\frac{\partial}{\partial\nu_l}
f(x,\nu_l,\epsilon)
\\
=&
-\epsilon
\frac{(-1)^{N}}{f(x,\nu_l,\epsilon)}
\frac{\partial^{N}}{\partial x^{N}}
f(x,\nu_l,\epsilon)
+\nu_l
\frac{(-1)^{N-1}}{f(x,\nu_l,\epsilon)}
\frac{\partial^{N-1}}{\partial x^{N-1}}
f(x,\nu_l,\epsilon)
\nonumber
\\
=&
-
\frac{
\epsilon}{
(1-\epsilon^2)^{N/2}}
H_N
\left(
\frac{x-\epsilon\nu_l}{\sqrt{1-\epsilon^2}}
\right)
+
\frac{
\nu_l}{
(1-\epsilon^2)^{(N-1)/2}}
H_{N-1}
\left(
\frac{x-\epsilon\nu_l}{\sqrt{1-\epsilon^2}}
\right)
\approx \nu_l H_{N-1}(x) \;.
\end{align}
We thus obtain
\be
\frac{1}{(N-1)!}\left\langle\nu_s^{N-1}\nu_l\right\rangle_c 
H_{N-1,1}(x,\nu_l,\epsilon)\approx \frac{\nu_l}{(N-1)!}
\left\langle\nu_s^{N-1}\nu_l\right\rangle_c H_{N-1}(x) 
+{\cal O}(\nu_l^2)\;.
\ee
On expanding the numerator of Eq.(\ref{eq:dpkng1}) at first order in 
$\nu_l$, we can isolate the Gaussian contribution, which is
\be
\frac{\frac{d}{dM}\int_\nu^\infty\!\!dx\,\epsilon\nu_l x\,e^{-x^2/2}}
{\frac{d}{dM}\int_\nu^\infty\!\!dx\,e^{-x^2/2}}=
\epsilon\nu_l\left(\nu-\frac{1}{\nu}\right)=
\left(\frac{\s_{0\times}^2}{\s_{0l}^2}\right)\,b_1\,\d_l \;,
\ee
in agreement with the linear PBS bias for the Press-Schechter mass function
derived in \refsec{MLB}.  We now retain all the terms linear in the
higher-order cumulants ($N\geq 3)$ in the linear expansion of 
\refeq{P1hat} and \refeq{P2hat}) and obtain
\begin{align}
\label{eq:dpkng2}
\d_h(\d_l) & \approx
\left(\frac{\s_{0\times}^2}{\s_{0l}^2}\right)\,b_1\,\d_l
+\left(\frac{\s_{0\times}^2}{\s_{0l}^2}\right)\,b_1\,e^{\nu^2/2}
\Biggl[\sum_{N=3}^\infty\frac{d}{dM}\!\int_\nu^\infty\!\!dx\,
\frac{\left\<\nu_s^N\right\>_c}{N!}H_N(x)e^{-x^2/2}\Biggr]
\frac{dM}{d\nu}\d_l \\
&\quad -e^{\nu^2/2}
\Biggl[\sum_{N=3}^\infty\frac{d}{dM}\!\int_\nu^\infty\!\!dx\,
\epsilon\nu_l\frac{\left\<\nu_s^N\right\>_c}{N!}H_N(x)\,x\, e^{-x^2/2} 
\Biggr]\frac{dM}{d\nu}\nonumber \\
&\quad -e^{\nu^2/2}
\Biggl\{\sum_{N=3}^\infty\frac{d}{dM}\!\int_\nu^\infty\!\!dx\,
\biggl[-\frac{\left\<\nu_s^N\right\>_c}{(N-1)!}\epsilon\nu_l
+\frac{\left\<\nu_s^{N-1}\nu_l\right\>_c}{(N-1)!}\nu_l\biggr]
H_{N-1}(x)\,e^{-x^2/2}\Biggr\}\frac{dM}{d\nu}\nonumber\;.
\end{align}
Using the generating function $\exp(xt-t^2/2)=\sum_N H_N(x) t^N/N!$, 
we can easily evaluate the integrals over the Hermite polynomials. In 
particular, we find for $N\geq 2$:
\be
\int_\nu^\infty\!\!dx\,x H_N(x)\,e^{-x^2/2} = 
\Bigl[\nu H_{N-1}(\nu)+H_{N-2}(\nu)\Bigr]e^{-\nu^2/2} \;.
\ee
On inserting this expression into Eq.(\ref{eq:dpkng2}), taking the 
derivative with respect to $M$ and employing the recurrence relation 
$H_{N+1}(x)=x H_N(x)-N H_{N-1}(x)$, the conditional overabundance of 
halos simplifies to
\begin{align}
\label{eq:dpkng3}
\d_h(\d_l) & \approx
\left(\frac{\s_{0\times}^2}{\s_{0l}^2}\right)\,b_1\,\d_l
-\left(\frac{\s_{0\times}^2}{\s_{0l}^2}\right)\,b_1
\sum_{N=3}^\infty\frac{1}{N!} 
\Bigl[\s_{0s}^2\d_c^{-1}\left\<\nu_s^N\right\>_c' H_{N-1}(\nu)
+\left\<\nu^N_s\right\>_c H_N(\nu)\Bigr]\d_l \\
&\quad - \left(\frac{\s_{0\times}^2}{\s_{0l}^2}\right)
\sum_{N=3}^\infty\frac{1}{N!}
\biggl[-\s_{0s}^2\d_c^{-1}\left\<\nu_s^N\right\>_c' H_N(\nu)
+\left\<\nu_s^N\right\>_c
\left(-H_{N+1}(\nu)+\frac{H_N(\nu)}{\nu}
\right)\biggr]\frac{\d_l}{\s_{0s}} 
\nonumber \\
&\quad +\sum_{N=3}^\infty\frac{1}{(N-1)!}
\Bigl[\s_{0s}^2\d_c^{-1}\left\<\nu_s^{N-1}\nu_l\right\>_c'H_{N-2}(\nu)
+\left\<\nu_s^{N-1}\nu_l\right\>_c H_{N-1}(\nu)\Bigr]
\frac{\d_l}{\s_{0l}}\nonumber\;.
\end{align}
where a primed variable $X'$ now designates $\partial X/\partial\s_{0s}$
(we have used the fact that $d\nu=-\delta_cd\s_{0s}/\s_{0s}^2$). 

\subsection{Non-Gaussian bias corrections}

In order to calculate the non-Gaussian bias corrections, we have to 
compute the derivative of the $N$-point cumulants 
$\left\<\nu_s^N\right\>_c=\s_{0s}^{N-2}S_s^{(N)}$
and $\left\<\nu_s^{N-1}\nu_l\right\>_c\equiv
\Sigma_\times^2/\left(\s_{0s}^{N-1}\s_{0l}\right)$  with respect to 
$\s_{0s}$. These are
\begin{equation}
\left\<\nu_s^N\right\>_c'=\s_{0s}^{N-3}S_s^{(N)}
\biggl[\left(N-2\right)+\frac{\partial\ln S_s^{(N)}}
{\partial\ln\s_{0s}}\biggr]\;,
\qquad
\left\<\nu_s^{N-1}\nu_l\right\>_c'=\frac{1}{\s_{0s}^N\s_{0l}}
\biggl[\s_{0s}\frac{\partial(\Sigma_\times^2)}{\partial\s_{0s}}
-\left(N-1\right)\Sigma_\times^2\biggr]\;.
\end{equation}
Replacing the Hermite polynomials with the Gaussian peak-background
split biases inferred from the Press-Schechter multiplicity function
[\refeq{bNdef}],
\be
\label{eq:pbias}
b_N(\nu)
=\frac{1}{\s_{0s}^N}\frac{H_{N+1}(\nu)}{\nu},
\ee
the conditional overabundance of halos can be recast into
\begin{align}
\label{eq:dpkng4}
\nonumber
\d_h(\d_l) & \approx
\left(\frac{\s_{0\times}^2}{\s_{0l}^2}\right)\,b_1\,\d_l
-\left(\frac{\s_{0\times}^2}{\s_{0l}^2}\right)
\sum_{N=3}^\infty\frac{S_s^{(N)}}{N!}
\biggl[\left(N-2\right)+\frac{\partial\ln S_s^{(N)}}
{\partial\ln\s_{0s}}\biggr]
\s_{0s}^{2(N-2)}
\Bigl(b_1 b_{N-2}-b_{N-1}\Bigr)\d_l
\\
&\quad -\left(\frac{\s_{0\times}^2}{\s_{0l}^2}\right)
\sum_{N=3}^\infty
\frac{S_s^{(N)}}{N!}
\s_{0s}^{2(N-2)}
\left(
\delta_cb_1b_{N-1} + b_{N-1} - \delta_c b_N
\right)
\d_l \nonumber \\
&\quad +\sum_{N=3}^\infty \frac{\s_{0s}^{-2}}{(N-1)!}
\Biggl[\left(\frac{(\Sigma_\times^2)'}{\s_{0l}^2}\right)
\s_{0s} b_{N-3}-\left(\frac{\Sigma_\times^2}{\s_{0l}^2}\right)
\left(N-1\right) b_{N-3}+\left(\frac{\Sigma_\times^2}
{\s_{0l}^2}\right) \d_c b_{N-2}\Biggr] \d_l
\;.
\end{align}
We can now read off the
scale-independent correction $\D\bias{I}^{(\iota)}$ (involving the terms 
proportional to $\s_{0\times}^2/\s_{0l}^2$) and a scale-dependent correction 
$\D\bias{I}^{(\kappa)}$ (involving the terms $\Sigma_\times^2/\s_{0l}^2$ 
and $(\Sigma_\times^2)'/\s_{0l}^2$) to the first order Gaussian bias $b_1$. 
The non-Gaussian bias  contribution thus is 
$\D\bias{I}=\D\bias{I}^{(\iota)}+\D\bias{I}^{(\kappa)}$. After some 
manipulation, the scale-independent non-Gaussian bias correction reads
\be
\label{eq:dbi}
\D\bias{I}^{(\iota)}(R_s,X) = 
-\frac{S_s^{(N)}\!(X)}{N!}\s_{0s}^{2(N-2)}
\left\{
\biggl[\left(N-2\right)+\frac{\partial\ln S_s^{(N)}(X)}
{\partial\ln\s_{0s}}\biggr]\bigl(b_1 b_{N-2}- b_{N-1}\bigr) 
-
\left(
\delta_cb_1b_{N-1} + b_{N-1} - \delta_c b_N
\right)
\right\}\;,
\ee
where $X$ is again a vector of variables describing the amplitude and shape of
the primordial $N$-point function. 
In order to write down an explicit expression for the scale-dependent, 
non-Gaussian bias correction, we use the definition of the kernel
${\cal S}(k,R_s)$ [\refeq{Skernel}], yielding
\be
\Sigma_\times^2
= 4 \s_{0s}^2 
\int\!\!\frac{d^3k}{(2\pi)^3} \Pp(k) 
\Mm_l(k)
{\cal F}_s^{(N)}(k,X)\;.
\ee 
Then, by definition of the linear halo bias, $\d_h(\vk) = b(k) \d_s(\vk)$,
correlating the last line of \refeq{dpkng4} with $\delta_l$ yields
\ba
\nonumber
&\int\!\!\frac{d^3k}{(2\pi)^3} \,
\Delta b_I^{(\kappa)}\Pp(k)\Mm_s(k) \Mm_l(k)
\\
=&
\sum_{N=3}^\infty \frac{\s_{0s}^{-2}}{(N-1)!}
\biggl\{
(\Sigma_\times^2)' \s_{0s} b_{N-3}
-\Sigma_\times^2 
\Bigl[\left(N-1\right) b_{N-3}- \d_c b_{N-2}\Bigr]
\Biggr\} 
\nonumber
\\
=&
\int\!\!\frac{d^3k}{(2\pi)^3} \Pp(k) \Mm_l(k)\,
\sum_{N=3}^\infty \frac{4}{(N-1)!}
\Biggl\{
\frac{d (\s_{0s}^2{\cal F}_s^{(N)})}{d\s_{0s}} \s_{0s}^{-1} b_{N-3}
-
{\cal F}_s^{(N)}
\Bigl[ \left(N-1\right) b_{N-3}- \d_c b_{N-2}\Bigr]
\Biggr\}\;,
\ea
from which we can read off the scale-dependent non-Gaussian bias correction
as
\be
\label{eq:dbk}
\D\bias{I}^{(\kappa)}(k,R_s,X) = \frac{4}{(N-1)!}\,
\Bigg\{b_{N-2}\,\delta_c + b_{N-3} \left[3-N + 
\frac{\partial\ln {\cal F}_s^{(N)}\!(k,X)  }{\partial\ln\s_{0s}}\right ]
\Bigg\} {\cal F}_s^{(N)}\!(k,X) {\cal M}_s^{-1}(k)\;. 
\ee
This is the main result of this Section. 
In the high-peak limit, $b_{N-2}\gg b_{N-3}$ and the first term in the 
curly bracket dominates. Therefore, we exactly recover the results of 
\cite{2008ApJ...677L..77M,2010PhRvD..81b3006D,2009ApJ...706L..91V,
2010arXiv1010.3722S} for the constant $\fnl$, constant $\gnl$, folded 
shape and $k$-dependent $\fnl$, respectively. The second term in the 
curly brackets arises owing to the mass-dependence of the reduced 
cumulants $S_s^{(N)}$. As we will see shortly, this term agrees with 
the correction derived in \refsec{GPBS} in the limit $k\to 0$.

Note that \cite{2010arXiv1012.1859C} also employed the bivariate Edgeworth 
expansion to explore the effect of a local primordial trispectrum on the 
(configuration space) bias of tracers. However, they did not derive any 
explicit expression for the non-Gaussian bias.

\begin{figure*}
\center
\resizebox{0.49\textwidth}{!}{\includegraphics{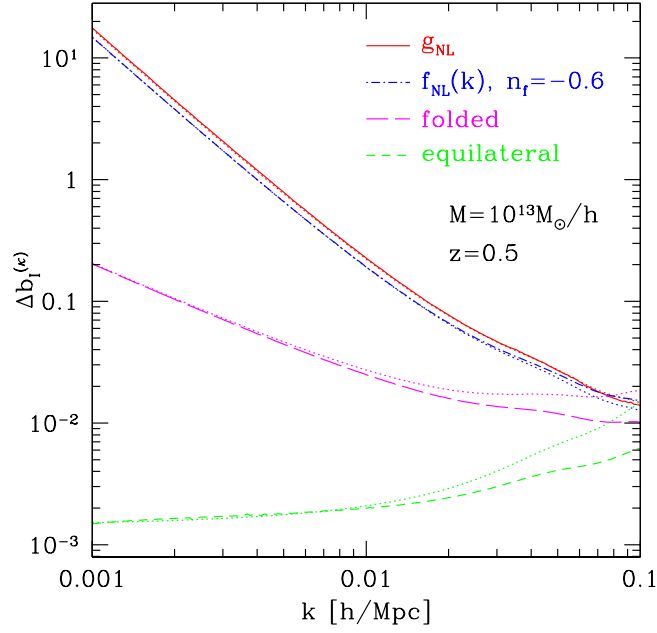}}
\caption{A comparison between the non-Gaussian scale-dependent bias 
correction \refeq{dbk} and its low-$k$ limit \refeq{Dbo-general} for 
some of the bispectrum shapes and the local trispectrum considered in 
this work. In all cases, a dotted curve represents the low-$k$ limit. 
Results are shown as a function of $k$ for halos of mass 
$M=5.3\times 10^{13}\hmsun$ at $z=0.5$, assuming $\fnl=100$ (for the 
bispectra) and $\gnl=10^6$ (for the local trispectrum).}
\label{fig:dbk}
\end{figure*}

\subsection{Comparison of the PBS approaches}
\label{sec:compare}

Interestingly, if we ignore the mass-dependence of the cumulants, then the 
$k$-dependence of \refeq{dbk} is exactly the same as that predicted by the 
correlation of thresholded regions (see \refsec{MLB}).  This follows from
expanding the non-Gaussian density field in cumulants, which is also done 
in the expansion of the correlation function of thresholded regions.  
By contrast, our first formulation of the peak-background split (see 
\refsec{GPBS}) leads to a different $k$-dependence on smaller scales.  
This difference arises because we have assumed that the long-wavelength 
perturbation is constant over some intermediate scale $R$ over which the 
halo abundance is averaged. This is a sensible assumption as long as the 
scale ``$L$'' over which we measure the clustering of halos is much larger 
than the scales that contribute to $\s_{0s}$. Then, the kernel $\o$ in 
\refeq{sigmao} is indeed evaluated in the squeezed limit, $k_s \gg k$, and 
both PBS formulations agree exactly. 
To see this explicitly, we write \refeq{dbk} for the cases of $N=3$ and 
$N=4$:
\ba
\D\bias{I}^{(\kappa)}(k,R_s,N=3) =\:& 2
\Bigg\{b_1\,\delta_c + \frac{\partial\ln {\cal F}_s^{(3)}\!(k)}
{\partial\ln\s_{0s}}
\Bigg\} {\cal F}_s^{(3)}\!(k) {\cal M}_s^{-1}(k)\label{eq:b30}\\
\D\bias{I}^{(\kappa)}(k,R_s,N=4) =\:& \frac{4}{6}
\Bigg\{b_2\,\delta_c + b_1\left[1+\frac{\partial\ln {\cal F}_s^{(4)}\!(k)}
{\partial\ln\s_{0s}}\right ]
\Bigg\} {\cal F}_s^{(4)}\!(k) {\cal M}_s^{-1}(k).\label{eq:b40}
\ea
In the large-scale limit, we can use the same approximations made in 
\refsec{GPBS}, i.e. assume that $k$ is much smaller than the scales which 
contribute significantly to the integrand in \refeq{fshape}. On inserting 
the definition of the kernels $\wt,\;\wth$ [\refeqs{wt}{wth}], we obtain
\ba
{\cal F}_s^{(3)}\!(k) \stackrel{k\to 0}{=}\:& \fnl \frac{\s_{\o s}^2(k)}
{\s_{0s}^2}\\
{\cal F}_s^{(4)}\!(k) \stackrel{k\to 0}{=}\:& \frac{3}{4} \gnl \s_{0s}^2 
\Sk_{\o s}(k).
\ea
Substituting these expressions into \refeqs{b30}{b40}, we eventually 
recover \refeq{Dbo-general} in \refsec{GPBS}.  

On smaller scales $k\gtrsim 0.02\iMpch$ around which the matter power 
spectrum peaks, the separation of scales ``$L$'' and ``$S$'' is no longer 
accurate and the predictions of \refeq{dbk} diverge from the $k\to 0$ limit.  
In Fig. \ref{fig:dbk}, the exact $k$-dependence of the non-Gaussian bias 
correction predicted by the correlated PBS approach, \refeq{dbk}, is 
compared to that predicted by the low-$k$ expression, \refeq{Dbo-general}. 
We can see that the latter is accurate to a few percent at wavenumber 
$k\lesssim 0.01\hmmpc$. Only for the folded and equilateral shape does
the low-$k$ expression yield a noticeably larger non-Gaussian bias 
correction on scales $k\gtrsim 0.01\hmmpc$. This is also true for the 
orthogonal template (not shown in the figure since it is essentially equal 
to the folded case). The exact difference, however, depends somewhat on 
halo mass and redshift.  
A quantitative comparison of the scale-dependent bias predicted by the 
uncorrelated PBS approach with that obtained from the statistics of 
thresholded regions can also be found in \cite{fsmk} (note however that 
the new term derived in this work is not included there).  

Finally, while in the limit $k\to 0$ \refeq{dbk} reproduces the well-known 
result for the local scale-independent $\fnl$ model
\cite{2008PhRvD..77l3514D,2008ApJ...677L..77M}, at finite $k$ this 
expression receives a negative correction from the second term proportional
to $\partial\ln{\cal F}_s^{(3)}/\partial\ln\s_{0s}$ that increases with 
wavenumber.  At $k=0.05\hmmpc$ for instance, the suppression is $\sim$1\% 
and $\sim$4\% for biased tracers with $b_1^{\rm E}\sim 2$ and 3.5, 
respectively.  

\section{Peak-background split vs Thresholding}
\label{sec:PBS-thr}

We now compare our final result \refeq{dbk}, with the result from 
thresholding in the high-peak limit,
\be
\Delta \bo^{\rm (hp)}(k)
= 
\frac{4 b_{N-1}}{(N-1)!} \s_{0s}^2
\frac{{\cal F}_s^{(N)}(k)}{\Mm_s(k)}\;,
\label{eq:dbk-hp}
\ee
obtained from \refeq{dbk-hp1} by replacing $c_N$ with  $b_N$. 
We see two important differences.  Firstly, in the thresholding 
approach (which is equivalent to local biasing), the correction to 
the halo power spectrum induced by a primordial $N$-point function 
is proportional to $b_{N-1}$.  In the 
PBS approach on the other hand, the correction comes in through 
the dependence of the halo mass function on the $(N-1)$-th moment
$S_s^{(N-1)}$ of the small-scale density field. The latter is 
proportional to $b_{N-2}$ when the Edgeworth approximation method is
applied to the halo mass function. The simulation results for all
types of primordial non-Gaussianity simulated so far clearly follow 
the dependence on $b_{N-2}$ rather than $b_{N-1}$, thus favoring the 
interpretation provided by the PBS approach.  

Secondly, the term proportional to 
$\partial\ln{\cal F}_s^{(3)}/\partial\ln\s_{0s}$ in the PBS prediction
[\refeq{dbk}] is absent in the thresholding approach. In 
\cite{2011PhRvD..84f1301D}, 
we show that the inclusion of this term yields a good match to the 
simulated halo bias in non-Gaussian models beyond the simplest, local 
quadratic non-Gaussianity with scale-independent $\fnl$. 
In the thresholding approach on the other hand, we associate the 
correlation of regions above a threshold $\d_c(z)$ in the linear density 
field smoothed at a \emph{fixed} scale $R_s$ with that of halos above 
a mass threshold $M(R_s)$ at redshift $z$. However, 
halos spanning some mass interval should be identified with Lagrangian 
regions spanning a \emph{range} of smoothing scales. 
Consequently, the abundance of halos in a mass interval $[M,M+dM]$ not 
only depends on the cumulants $S_s^{(N-1)}$ of the density field smoothed on 
scale $R_s$, but also on the variation of these cumulants with $R_s$ 
(parameterized through $\partial\ln S_s^{(N-1)}/\partial\ln\s_{0s}$).  

An alternative way of seeing this is to describe the abundance of halos
in a non-Gaussian density field through an effective significance
$\hat\nu(\nu,S_s^{(m)})$, which is defined upon requiring 
\be
\nhh = \frac{\rhob}{M^2} f_{\rm NG}(\nu) \frac{d\ln\nu}{d\ln M} 
= \frac{\rhob}{M^2} f_{\rm G}(\hat\nu) \frac{d\ln\hat\nu}{d\ln M}\;.
\label{eq:nhNG}
\ee
In the case of a Press-Schechter mass function [\refeq{P1hat}], 
$\hat\nu$ is given by
\be
\hat\nu = \nu\left(1 - \sum_{m=3}^\infty
\frac{1}{m!} S_s^{(m)}\: \s_{0s}^{2(m-2)} b_{m-2}\right)
\ee
The Jacobian $d\ln\hat\nu/d\ln M$ in \refeq{nhNG} involves 
$\partial\ln S_s^{(m)}/\partial\ln\nu$, showing that $\nhh$
depends on the scale-dependence of the cumulants.


\section{Conclusion}
\label{sec:conclusion}

We have carefully re-examined the derivation of the effect of primordial
non-Gaussianity on the large-scale clustering of tracers (such as galaxies
and clusters) beyond the local $\fnl$ model, using the statistics of 
thresholded regions as well as two formulations of
the peak-background split (PBS).  We have shown that the thresholding 
approach is equivalent to local biasing, when considering the
leading order contributions from non-Gaussianity.  This approach
predicts the same scale-dependence as the PBS approach in the limit
$k\to 0$, in agreement with the findings of \cite{fsmk}. However, unless we
consider the high-peak limit (which is never attained by real tracers), PBS 
and thresholding predict different amplitudes for the non-Gaussian 
contribution to the linear bias. 

We have presented two complementary peak-background split derivations of the 
effect of non-Gaussianity. In the first approach, the separation of scales
is invoked to split the {\it Gaussian} density field into uncorrelated short- 
and long-wavelength perturbations.  This allows us to isolate the 
mode-coupling effect responsible for the scale-dependent bias induced by
non-Gaussianity.  In the second approach, the
separation of scales is invoked to expand the ratio of the unconditional to
conditional mass function in terms of large-scale perturbations in the
{\it non-Gaussian} density field. Notice that no assumption of separation of 
scales is made in the thresholding approach, where biasing is a function of 
the local density only.
While in the second PBS approach we have restricted ourselves to the
case of a Press-Schechter mass function, we have nonetheless been able to 
identify the non-Gaussian bias correction to the linear bias. Both PBS
approaches predict exactly the same correction in the limit $k\to 0$ (once
the Press-Schechter expressions for the Gaussian biases are identified with
$b_N$).  
While they depart at higher wavenumbers ($k\gtrsim 0.02\iMpch$), 
this deviation is not very significant for the local 
or folded type of non-Gaussianity where the non-Gaussian bias correction is 
strongly suppressed at small scales.

In both approaches, we uncover a new term depending on the scale-dependence
of the small-scale moments of the density field induced by non-Gaussianity.  
Physically, this term is induced by the mapping from local significance
$\nu = \d_c/\s_{0s}$ to mass $M$:  a scale-dependent modulation of
$\s_{0s}$ changes the interval $d\nu$ corresponding to a fixed mass interval
$dM$.   
This correction to the high-peak expression of the linear non-Gaussian bias 
has not been pointed out in any previous work.  It can be very large for all 
the models considered here, except for the local bispectrum with constant 
(i.e., $k$-independent) $\fnl$.  Moreover, we have found very good overall 
agreement between the PBS predictions and the simulated non-Gaussian halo 
bias \cite{2010PhRvD..81b3006D,2010arXiv1010.3722S,2011arXiv1102.3229W}
for the local $\gnl\phi^3$ model, the local $\fnl\phi^2$ model with 
$k$-dependent $\fnl$, and the orthogonal bispectrum.  This comparison is 
detailed in a companion Letter to this paper \cite{2011PhRvD..84f1301D}. 
Consequently, 
the simulation results rule out thresholding, and more generally local 
biasing, as a viable approach to predicting the impact of primordial 
non-Gaussianity on halo clustering. These new accurate predictions can be 
combined with optimal weighting schemes 
\cite{2009JCAP...03..004S,2009PhRvL.103i1303S,2010PhRvD..82d3515H,
2010MNRAS.tmp.1878C,2011arXiv1104.2321H,2011arXiv1104.3862B} 
in order to extract information on the scale-dependent bias from numerical 
simulations and forthcoming galaxy surveys. 

In order to further test the PBS approach with numerical simulations, it 
will be important to take into account the scale-independent correction 
$\D\bias{I}^{(\iota)}$ induced by non-Gaussianity through its impact on the 
abundance of halos.  In the case of local cubic non-Gaussianity, it will also 
be necessary to measure the Gaussian second-order bias factor $b_2$ directly 
from the simulations, as the effect on the linear bias scales with $b_2$.  

Finally, a natural generalization of the conditional mass function approach 
discussed in \refsec{CPBS} is a derivation of the non-Gaussian bias factors
within the excursion set formalism, for generic moving barriers and 
non-Gaussian initial conditions
\cite{2009MNRAS.398.2143L,2010MNRAS.405.1244M,2010ApJ...717..526M,
2010arXiv1007.1903D}.
We leave these issues for more detailed future treatments.

\section*{Acknowledgements}

We are grateful to Tobias Baldauf, Olivier Dor\'e, Chris Hirata, Marc
Kamionkowski, Eichiiro Komatsu, Rom\'an Scoccim\'arro, Emiliano Sefusatti, 
Leonardo Senatore, Ravi Sheth for many fruitful discussions, and to Sirichai 
Chongchitnan and Ravi Sheth for comments on an early version of this 
manuscript. VD wishes to thank Theoretical Astrophysics at Caltech and the 
Center for Cosmological Physics at Berkeley for hospitality during the 
completion of parts of this work. DJ and FS are supported by the Gordon and 
Betty Moore Foundation at Caltech. VD is supported by the Swiss National 
Foundation under contract 200021-116696/1 and FK UZH 57184001.

\bibliography{PBSng}

\end{document}